\documentclass[aps,prd,preprint,superscriptaddress]{revtex4}

\usepackage{graphics}
\usepackage{epsfig}
\usepackage{latexsym}
\usepackage{colordvi}
\usepackage{amsmath}
\usepackage{amssymb}

\newcommand{\fun}{$F_2^{\gamma}(x,Q^2)$ }
\newcommand{\ra}{\rightarrow}
\newcommand{\gam}{^{\gamma}}
\newcommand{\fund}{$F_2^{\gamma}(x,Q^2)$}
\newcommand{\be}{\begin{equation}}
\newcommand{\ee}{\end{equation}}
\newcommand{\ba}{\begin{eqnarray}}
\newcommand{\ea}{\end{eqnarray}}
\newcommand{\etal}{{\it et al.}}

\def\z0{Z}
\def\eg{{e.g.} }
\def\ie{{i.e.} }

\begin{document}

\begin{flushright} 
CERN-TH/2002-362 \\
IFT-22/2002 \\
UG-FT-138/02 \\
CAFPE-8/02 \\
DESY 02-118
\end{flushright}


\title{
A New 5-Flavour LO Analysis and Parametrization\\
 of Parton Distributions in the Real Photon }

\author{F.~Cornet}
\affiliation{Departamento de F\'{\i}sica Te\'orica y del Cosmos,
  Universidad de Granada, Campus de Fuente Nueva, E-18071, Granada,
  Spain}

\author{P.~Jankowski}
\affiliation{Institute of Theoretical Physics, Warsaw University,
 ul. Ho\.za 69, 00-681 Warsaw, Poland}

\author{M.~Krawczyk}
\affiliation{Institute of Theoretical Physics, Warsaw University,
 ul. Ho\.za 69, 00-681 Warsaw, Poland}
\affiliation{CERN, TH Division, CH-1211 Gen\`eve 23, Switzerland}

\author{A.~Lorca}
\affiliation{Departamento de F\'{\i}sica Te\'orica y del Cosmos,
  Universidad de Granada, Campus de Fuente Nueva, E-18071, Granada,
  Spain}
\affiliation{Deutsches Elektronen-Synchrotron, DESY Zeuthen, Platanenallee 6, D-15738 Zeuthen, Germany}


\begin{abstract}
New, radiatively generated, LO quark ($u,d,s,c,b$) and gluon densities in a 
real, unpolarized photon are presented. We perform a global 3-parameter fit, 
based on LO DGLAP evolution equations, to all available data for the structure 
function $F_2^{\gamma}(x,Q^2)$. We adopt  a new theoretical approach called 
ACOT($\chi$), originally introduced for the proton, to deal with the 
heavy-quark thresholds. This defines our basic model (CJKL model), which gives 
a very good description of the experimental data on $F_2^{\gamma}(x,Q^2)$, for
both $Q^2$ and $x$ dependences. For comparison we perform a standard fit using 
the Fixed Flavour-Number Scheme (FFNS$_{CJKL}$ model), updated with respect to
the previous fits of this type. We show the superiority of the  CJKL fit over 
the FFNS$_{CJKL}$ one and  other LO fits to the $F_2^{\gamma}(x,Q^2)$ data. 
The CJKL model gives also the best description of the LEP data on the $Q^2$ 
dependence of the $F_2\gam$, averaged over various $x$-regions, and the 
$F_{2,c}\gam$, which were not used directly in the fit. Finally, a simple 
analytic parametrization of the resulting parton densities obtained with the 
CJKL model is given.
\end{abstract}


\maketitle

\section{Introduction}

The photon structure function was recognized as an important quantity 
already in the early days of the Parton Model \cite{zerwas}. It has attracted 
even more attention since the seminal paper by Witten \cite{WITTEN}, which 
shows that $F_2^{\gamma}$ can serve as a unique test of QCD. This expectation 
was based on the fact that the (asymptotic) point-like solution of the $Q^2$ 
evolution equation, summing the leading QCD corrections, can be obtained for 
$F_2^{\gamma}$ without additional assumptions. Further studies showed the need 
of the hadronic, VMD-type, contribution to $F_2^{\gamma}$, and consequently the
need of an input, as for every other hadronic structure function.

The structure function $F_2^{\gamma}$ is extracted from measurements in deep 
inelastic scattering on a photon target, which can be performed in $e^+e^-$ 
experiments. The $F_2^{\gamma}$ data are used to construct parametrizations 
of the parton distributions in the photon. The need for a resolved photon 
interaction, \ie where a photon interacts via its partonic agents, has become 
apparent in other type of processes involving photons, namely in the 
production of particles with a large transverse momentum. A recent review of 
the experimental situation and the existing parametrizations can be found in 
\cite{NISIUS}.

Our motivation for a new global analysis of the $F_2^{\gamma}$ data and for 
constructing a new parton parametrization for a real unpolarized photon is 
twofold. On the one hand, there is a vast amount of new experimental data on 
$F_2^\gamma(x,Q^2)$ that has not been used yet to produce the parton 
parametrizations for the photon. Two recent parametrizations, GRV \cite{grv92}
and GRS \cite{grs}, used respectively about 70 and 130 experimental points, 
while at present a total of 208 independent \fun points exist. On the other 
hand, there are discrepancies between the theoretical calculations and 
experimental results for some processes initiated by real photons in which 
heavy quarks are produced\footnote{Discrepancies were also observed in 
{$b$-quark} production at the $p\bar p$ collider \cite{bb}.}. 
Let us just mention here the {$D^*$}- and $D_s$-meson photoproduction 
\cite{dstar}, \cite{ds} and \cite{Frix} or the $D^*$-meson production with 
associated dijets \cite{dstar} at HERA, as examples. The disagreement between 
the theoretical and experimental results is even more profound for the open 
beauty-quark production in both HERA \cite{bhera} and LEP \cite{bqOPAL,bqL3} 
measurements.

The idea of the radiatively generated parton distributions has been 
successfully introduced by the GRV group first to describe the parton 
distributions in the nucleon \cite{grvn} and pion \cite{grvp}, and later to 
create the LO and NLO parton parametrization for the real \cite{grv92} and 
virtual \cite{grst} photon. Here we follow this approach for a real photon 
case, limiting ourselves, to the analysis based on the LO QCD. The NLO 
analysis is under preparation.

As mentioned above there is a problem with the QCD description of heavy-quark 
production in processes initiated by photons. Therefore, our analysis 
especially focuses on the heavy-quark contributions to the \fund. We apply a 
new Variable Flavour-Number Scheme (VFNS) approach, denoted by ACOT($\chi$),
proposed for heavy-quark production in the $ep$ collision 
(``electroproduction'') in \cite{tung}. For comparison we perform a standard
Fixed Flavour-Number Scheme (FFNS) fit as well. Since these two approaches are 
based on very distinct schemes, and since they need different evolution 
programs, we will refer to them as to two models, CJKL (ACOT($\chi$) type) and 
FFNS$_{CJKL}$ models, respectively.

Our paper is divided into six parts. In section 2 we describe various
approaches including the ACOT($\chi$) scheme \cite{tung}, applied to the 
production of heavy quarks in hadronic processes. Section 3 is devoted to the
description of the $F_2^{\gamma}$ in LO QCD, paying special attention to an
implementation of the ACOT($\chi$) scheme in the calculation of the \fund. In 
section 4 a description of the two global fits performed by us is given. In 
particular, we present the solutions of the DGLAP evolution in both models. We
describe in detail the assumptions for the input parton densities. In the fifth
section of the paper, the results of the global fits are discussed and compared
with the data for the \fund, and for the $F_2^{\gamma}(x,Q^2)$ averaged over 
various $x$-regions. A comparison with LEP data for $F_{2,c}\gam$
is presented in section 5 as well. The summary of the paper and an outlook of
work in progress can be found in section 6. Finally, in the appendix 
we give a simple parametrization of the CJKL (LO) parton distributions.


\section{Various schemes for a description of heavy-quarks production:
  the proton-target case}

In this section we describe various schemes, which are used in the calculation 
of the heavy-quark production in hadronic processes. There exist two standard 
schemes. In the FFNS, the light quarks ($u,d$ and $s$) and the heavy
ones ($c$ and $b$) are treated on a different footing. The 
light ones are treated as being massless, and together with the gluons, are 
the only partons in the proton. The massive charm- and beauty-quarks are 
produced in the hard subprocesses: they can only appear in the final state
of the process. In the second scheme, the Zero-mass Variable Flavour-Number 
Scheme (ZVFNS), when the characteristic hard scale of the process is larger 
than some threshold associated with a heavy quark, this quark is also 
considered as a massless parton in the proton, in addition to the three light 
quarks. In this way, the number of different types of quarks (flavours) that 
we treat as partons in the proton increases with the scale of the process.

For the Deep Inelastic Scattering on the proton (DIS$_{ep}$), where the 
structure functions of the proton are measured, the condition for considering 
a heavy quark to be a parton of the proton is given by a simple (kinematic) 
threshold condition for the total energy in the $\gamma^* p$ collision $W$, 
namely $W>2m_h$ (from now on we will denote the heavy quarks by $h$). It 
defines the kinematically allowed region for the production of a 
heavy-quark pair. However, for the structure functions, e.g. $F_2^p(x,Q^2)$, 
and further for the parton densities $q^p(x,Q^2)$, not $W$ but the virtuality 
of the probing photon $Q^2$ is considered to be a natural scale. In the 
inclusive production of heavy quarks, their transverse momentum or mass is 
often taken as a characteristic (hard) scale $\mu$, where 
$\mu\gg \Lambda_{QCD}$, at which parton densities of the initial hadrons are 
probed. The two, massive and massless, approaches are considered to be 
reliable in different $\mu$ regions. The FFNS loses its descriptive power 
when $\mu \gg m_h$; on the other hand the ZVFNS does not seem appropriate if 
$\mu \approx m_h$. In order to achieve a prescription working in all hard 
scale regions, various schemes trying to combine the two approaches have been 
proposed. They have a generic name: Variable Flavour-Number Schemes (VFNSs). 
The first of such approaches was introduced by the ACOT group in \cite{acot}. 
Other groups, such as RT \cite{rth}, BMSN \cite{bmsn} or CSN \cite{csnc}, created 
their own versions of the VFNS\footnote{Reviews of the VFNS existing in the literature can be found in \cite{tung}, \cite{csnc} and \cite{csnb}.}.

\bigskip

Let us discuss some aspects of the VFNS in more detail. The VFNS introduce 
the notion of ``active quarks'', for which the condition $\mu>m_q$ is 
fulfilled. Such quarks can be treated as (massless) partons of the initial 
hadron(s). Light ($u,d$ and $s$) quarks are always active because for them 
$\mu \gg \Lambda_{QCD} \ge m_q$. The heavy-quark densities vanish for 
$\mu \leq m_h$, otherwise they differ from zero like in the ZVFNS. For
example, for the charm quark we see that at $\mu = m_c$ we turn from a Three 
Flavour-Number Scheme to a Four Flavour-Number one. If $N_f$ equals the
number of active (massless) quarks, we define the $N_f$-FNS as one
where the $N_f$ first quarks are treated as light and the 
remaining quarks as heavy.

In calculations based on the VFNS we take into account the sum of all 
contributions, which would be included separately in the ZVFNS and FFNS. Such 
procedure requires a proper subtraction of the double-counted contributions. 
Such contributions have the form of the large logarithms $\ln \mu^2$, and are 
already resummed in the density $q_h(x,\mu^2)$.

An important aspect of the VFNS is the behaviour of the heavy-quark 
contributions in the threshold region. Let us discuss as an example the 
production of heavy quarks in DIS$_{ep}$. As was already mentioned, a heavy
quark can be considered as a parton of the proton if the centre of mass energy of
the hard process is $W>2m_h$. However, if we use $\mu^2$ equal to $Q^2$, 
where $Q^2=W^2 \frac{x}{1-x}$, and impose the threshold condition on $\mu^2$, then, 
for any $Q^2$, it may happen (for small enough $x$) that $q_h(x,Q^2)=0$ in the 
kinematically allowed region, \ie for $W>2m_h$. On the other hand, non-zero 
heavy-quark densities may appear in the kinematically forbidden region in 
the ($x,Q^2$) plane. Moreover, such conditions can lead to a very steep or 
even non-continuous growth of the heavy-quark distributions at the threshold.

In general one should ensure that all the ZVFNSs and relevant subtraction 
terms smoothly vanish when $W\to 2m_h$. Then, the non-zero contributions
should give only those terms which arise in the FFNS approach, since this 
approach should reliably describe the region $W \approx 2m_h$. Different 
threshold conditions were used in different analyses; in particular, the ACOT 
group proposed to use a variable $\mu^2$ given by
\be
\mu^2 = \begin{cases} 
m_h^2 + cQ^2(1-m_h^2/Q^2)^n & \mathrm{for} \quad Q^2>m_h^2, \\
m_h^2 & \mathrm{for} \quad Q^2\leq m_h^2,
\end{cases}
\end{equation}
where $c=0.5$ and $n=2$ in \cite{acot}. Still, in their
 approach the heavy-quark densities satisfy the boundary 
condition at $Q^2 = m_h^2$
\be
q_h(x,\mu^2) \begin{cases}
=0 & \mathrm{for} \quad \mu^2 = m_h^2 \quad (Q^2\leq m_h^2), \\
\neq 0 & \mathrm{for} \quad \mu^2 > m_h^2 \quad (Q^2>m_h^2).
\end{cases}
\end{equation}

Recently, a purely kinematic solution of the threshold-behaviour problem has
been found, on which the so-called ACOT($\chi$) scheme \cite{tung} is based. A 
new variable, $\chi_h \equiv x(1+4m_h^2/Q^2)$, has been introduced 
to replace the Bjorken $x$ as an argument in the heavy-quark $h$ density 
in the ZVFNS contributions. More details can be found in section 3.

Although the above discussion was focused on the proton case, the problems 
with the proper treatment of the heavy-quark thresholds in the parton 
distribution are very similar for any target. In this paper we adopt 
the ACOT($\chi$) scheme to the real photon case for the very first time.


\section{Description of $F_2^{\gamma}$ in the ACOT($\chi$) scheme}

In this section we first recall the basic facts related to the structure 
function $F_2^{\gamma}(x,Q^2)$ for the real photon. Then we introduce the 
ACOT($\chi$) approach for the photonic case.


\subsection{The parton densities in the photon}

The Deep Inelastic Scattering on a real photon (DIS$_{e\gamma}$)
allows us to measure the structure function $F_2^{\gamma}$, and also other structure 
functions, $F_1^{\gamma}$, $F_L^{\gamma}$, $\cdots$, via the process
\be
\gamma^* \gamma \ra \mathrm{hadrons},
\end{equation}
see [1--3]. In the Parton Model this is described at lowest order by 
the Bethe--Heitler (BH) process, $\gamma^* \gamma \to q\bar q$ 
(see Fig. \ref{BH}).

In the leading logarithmic ($\ln Q^2$) approximation or, in short, in the 
leading order of QCD (LO QCD), the photon structure function \fun can be 
written in terms of quark (antiquark) densities $q_i\gam (\bar q_i^{\gamma})$ 
as follows
\be
\frac{1}{x} F_2^{\gamma}(x,Q^2) = \sum_{i=1}^{N_f} e_i^2
(q_i^{\gamma} + \bar q_i^{\gamma})(x,Q^2),
\label{LOQCD}
\end{equation}
where $N_f$ is the number of different quark flavours, that can appear in the
photon (``active quarks''). Note that 
$q_i^{\gamma}(x,Q^2)=\bar q_i^{\gamma}(x,Q^2)$.

The evolution of the parton densities with $\ln Q^2$ is governed by 
the inhomogeneous DGLAP equations. In LO we have for a quark (similarly
for an antiquark) and a gluon density 
\ba
\frac{dq_i\gam(x,Q^2)}{d\ln Q^2}
&=& \frac{\alpha}{2\pi}e_i^2 k(x) + \frac{\alpha_s(Q^2)}{2\pi}
\int_x^1\frac{dy}{y} \Bigg[
 P_{qq}\left(\frac{x}{y}\right)q_i\gam(y,Q^2) + P_{qG}\left(\frac{x}{y}\right)G\gam(y,Q^2)\Bigg],
\label{DGLAP1}\\
\frac{dG\gam(x,Q^2)}{d\ln Q^2}
&=& \frac{\alpha_s(Q^2)}{2\pi}\int_x^1\frac{dy}{y} \Bigg[ 
P_{Gq}\left(\frac{x}{y}\right)\sum_{i=1}^{N_f} (q_i\gam+\bar q_i\gam)(y,Q^2) 
+ P_{GG}\left(\frac{x}{y}\right)G\gam(y,Q^2)\Bigg].
\label{DGLAP2}
\ea

The $k(x)$ term on the right-hand side of Eq. (\ref{DGLAP1}) comes from the Bethe--Heitler process of Fig. \ref{BH}; for 3 colours we have
\be
k(x) = 3\left[x^2+(1-x)^2 \right].
\label{kBethe}
\end{equation}
The functions $P_i(x)$ are the LO splitting functions \cite{ap}
\be
\begin{split}
P_{qq}(x) &= \frac{4}{3}\left[ \frac{1+x^2}{(1-x)_+} + \frac{3}{2}\delta(1-x)\right],\\
P_{qG}(x) &= \frac{1}{2}\left[x^2+(1-x)^2\right],\\
P_{Gq}(x) &= \frac{4}{3}~\frac{1+(1-x)^2}{x},\\
P_{GG}(x) &= 6\left[ \frac{x}{(1-x)_+}+\frac{1-x}{x}+x(1-x)\right]
+ \left[ \frac{11}{2}-\frac{N_f}{3} \right]\delta(1-x).
\end{split}
\end{equation}
Note that the function $k(x)$ describes  a photon into quark splitting, 
so one has 
$k(x)\equiv P_{q\gamma}(x)$.


\subsubsection{Heavy-quark contributions to $F_2\gam$ in the FFNS} 

\begin{figure}
\includegraphics[scale=0.7]{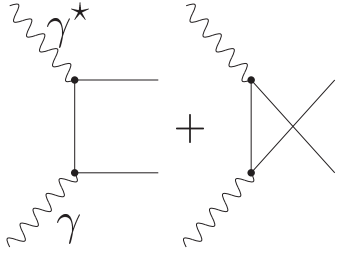}%
\caption{The Bethe--Heitler process $\gamma^* \gamma \rightarrow q \bar q$.\label{BH}}
\end{figure}

The standard FFNS approach corresponds to a number of ``active quarks'' 
($N_f$) =3, so only the light quarks (and their antiquarks) are taken
into account in the sum in Eq. (\ref{LOQCD}). The main heavy-quark 
contributions to $F_2^{\gamma}$ are obtained in this scheme from the 
corresponding Bethe--Heitler process (Fig. 1 with $q \ra h$)
\be
\gamma^{\star}\gamma \to h\bar h,
\end{equation}
keeping the heavy-quark masses in the calculation. It reads
\be
\frac{1}{x}F_{2,h}^{\gamma}(x,Q^2)|_{\mathrm{dir}} = 3\frac{\alpha}{\pi}e_h^4
\omega(x,Q^2),
\label{bhform}
\end{equation}
with
\ba
\omega(x,Q^2) &=& 
\beta \bigg[ -1+8x(1-x)-x(1-x)\frac{4m_h^2}{Q^2} \bigg] \nonumber\\
&&+\ln
\left(\frac{1+\beta}{1-\beta}\right)\bigg[x^2+(1-x)^2
+x(1-3x)\frac{4m_h^2}{Q^2}-x^2\frac{8m_h^4}{Q^4}\bigg] \\
\beta &=& \sqrt{1-\frac{4m_h^2x}{(1-x)Q^2}} = \sqrt{1-\frac{4m_h^2}{W^2}}. 
\label{bhformbis}
\ea
We call this contribution $F_{2,h}^{\gamma}|_{\mathrm{dir}}$ since here the real 
photon (\ie the target photon) interacts directly. However, there exists 
another heavy-quark contribution, related this time to the process with the 
resolved initial photon, namely 
\be
\gamma^{\star} G \gam \to h\bar h,
\end{equation}
with a gluonic parton of the photon target (as in Fig. \ref{BH} with 
$\gamma\to G^{\gamma}$); it gives
\be
F_{2,h}^{\gamma}(x,Q^2)|_{\mathrm{res}} = \frac{\alpha_s(Q^2)}{2\pi}e_h^2 
\int_{\chi_h}^1 \! \frac{x}{z} \omega \! \left( \frac{x}{z},Q^2\right) G\gam (z,Q^2) dz,
\label{res}
\end{equation}
where 
\be 
\chi_h \equiv x\left(1+\frac{4m_h^2}{Q^2}\right).
\label{defchi}
\end{equation}


\subsubsection{Heavy-quark contributions to $F_2\gam$ in the ZVFNS}

In the ZVFNS, the number of ``active quarks'' changes with the hard scale, as 
described in the previous section. For low scales the sum in Eq. (\ref{LOQCD})
extends to $N_f = 3$ but whenever a heavy quark threshold is surpassed the 
value of $N_f$ is increased by 1. It is worth mentioning that in some parton
parametrizations for a real photon the heavy-quark densities do appear; however
they are described in the threshold region $W \approx 2m_h$ only by the above 
Bethe--Heitler formula (Eqs. (10) and (11)). Moreover, instead of the restriction on 
$W$ one sometimes takes a (reasonable) condition on $Q^2$ (see, for instance, 
the GS parametrization ~\cite{gs}). Of course well above a heavy-quark 
threshold, such a quark can be included among the active (massless) quarks and 
then $N_f \to N_f+1$, see \eg \cite{grvp}.


\subsection{ACOT($\chi$) scheme for $F_2^{\gamma}$}

The ACOT($\chi$) prescription combines the FFNS and ZVFNS, so that we have 
to add all relevant contributions from both approaches. For the light-quark 
contributions we take the form given in Eq. (\ref{LOQCD}) with $N_f = 3$,
while for the heavy quarks we include the following terms:
\ba
\tilde F_2^{\gamma}(x,Q^2)|_{c,b} &=& \sum_{h(=c,b)}^2 \big[
x e_h^2 (q_h\gam+\bar q_h\gam)(x,Q^2) 
+ F_{2,h}^{\gamma}(x,Q^2)|_{\mathrm{dir}} + F_{2,h}^{\gamma}(x,Q^2)|_{\mathrm{res}}
\big],\qquad
\label{sum}
\ea
where $F_{2,h}^{\gamma}(x,Q^2)|_{\mathrm{dir}}$ and 
$F_{2,h}^{\gamma}(x,Q^2)|_{\mathrm{res}}$ are given in Eqs. (\ref{bhform}) and 
(\ref{res}), respectively.

In Eq. (\ref{sum}) we double-count some heavy-quark contributions. Indeed,
part of the $F_{2,h}^{\gamma}|_{\mathrm{dir}}$ contribution from 
$\gamma^{\star}\gamma \to h\bar h$ corresponds to the collinear 
configuration. Such a configuration leads to a contribution proportional to 
$\ln Q^2$ and is already included in the DGLAP equation for $q\gam_h(x,Q^2)$,
via the $k(x)$ term. Therefore we must subtract from (\ref{sum}) the following
terms 
\begin{equation}
F_{2,h}^{\gamma}|_{\mathrm{dir},\mathrm{subtr}}= x \ln \frac{Q^2}{m_h^2} 
3e_h^4 \frac{\alpha}{\pi}\left(x^2+(1-x)^2\right),
\end{equation}
coming from an exact solution of a part of the DGLAP equation, namely
\be
\frac{dq_h\gam (x,Q^2)}{d\ln Q^2} = \frac{\alpha}{2\pi}e_h^2 k(x),
\label{ss18}
\end{equation}
integrated over the $Q^2$ from $m_h^2$ to $Q^2$.

Similarly, $F_{2,h}^{\gamma}|_{\mathrm{res}}$ from the 
$\gamma^{\star}G^{\gamma} \to h\bar h$ process, has a $\ln Q^2$ part 
that corresponds to the collinear configuration already included
in the DGLAP equation for $q\gam_h(x,Q^2)$, via the 
$P_{qG}\left(\frac{x}{y}\right)G\gam(y,Q^2)$ term. The term to be subtracted reads, in 
this case:
\be
F_{2,h}^{\gamma}|_{\mathrm{res},\mathrm{subtr}}=\: 
x \ln \left(  \frac{Q^2}{m_h^2} \right) e_h^2 \frac{\alpha_s(Q^2)}{\pi} 
\int_{x}^1  \frac{dy}{y} P_{qG} \left( \frac{x}{y} \right) G\gam(y,Q^2).
\label{ss19}
\end{equation}

\begin{figure}
\includegraphics[scale=0.6]{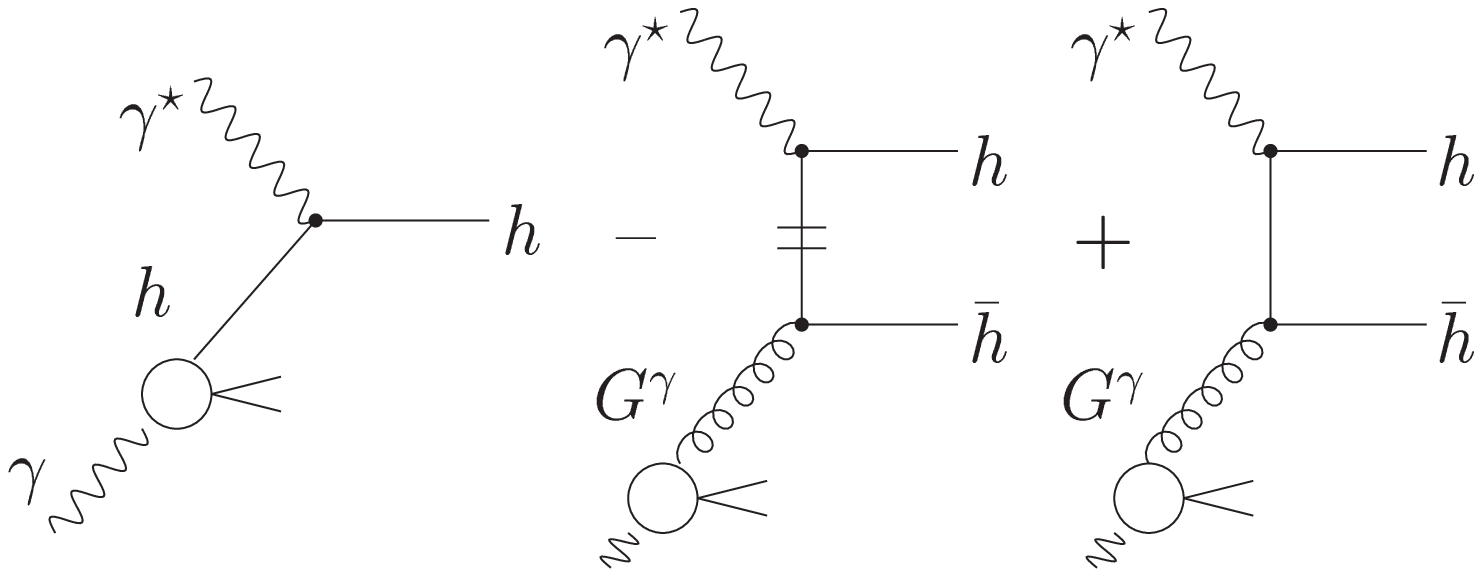}\\
\includegraphics[scale=0.6]{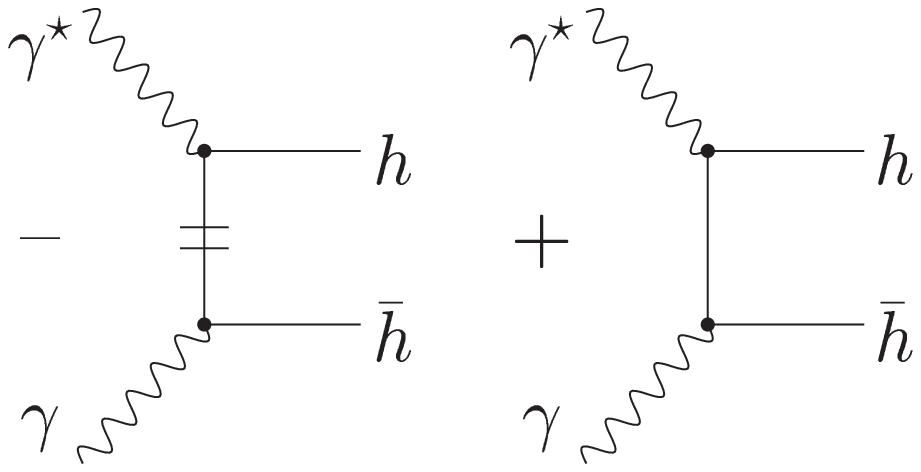}%
\caption{Graphical representation of the ACOT model for $F_2^{\gamma}$. The 
first diagram represents the ZVFNS contribution. The third (fifth) diagram 
shows the FFNS contribution of the resolved (direct) photon, while the second 
(fourth) diagram is the corresponding subtraction term.
\label{tunggraph}}
\end{figure}

It is based on an approximated solution for the other part of Eq. (\ref{DGLAP1}), 
namely 
\be
\frac{dq_h\gam (x,Q^2)}{d\ln Q^2} = \frac{\alpha_s(Q^2)}{2\pi}
\int_x^1\frac{dy}{y}\left[ P_{qG}\left(\frac{x}{y}\right)G\gam (y,Q^2) \right].
\label{ss20}
\end{equation}
The solution (\ref{ss19}) is obtained by the integration of Eq. (\ref{ss20}) 
over the same $Q^2$ region as above, after neglecting the $Q^2$ 
dependence\footnote{ The $Q^2$-dependence will appear back in the final 
solution, in both $\alpha_s$ and $G\gam$, where it is then a correction of higher 
order in $\alpha_s$.} of $\alpha_s$ and $G\gam$. The final subtraction term is
\begin{equation}
F_2^{\gamma}(x,Q^2)|_{\mathrm{subtr};c,b} = \! \sum_{h(=c,b)}^2 \! \big[
F_{2,h}^{\gamma}|_{\mathrm{dir},\mathrm{subtr}} + F_{2,h}^{\gamma}|_{\mathrm{res},\mathrm{subtr}} \big].
\label{sub}
\end{equation}
So, finally we have, for heavy-quarks: $F_2^{\gamma}(x,Q^2)|_{c,b}=\tilde F_2^{\gamma}(x,Q^2)|_{c,b}-F_2^{\gamma}(x,Q^2)|_{\mathrm{subtr};c,b}$. A graphical 
representation of all terms included in the analysis, Eqs. (\ref{sum}) and 
(\ref{sub}), is presented in Fig. \ref{tunggraph}.

Further, we need to ensure that all terms containing the heavy-quark $h$ 
disappear when $W \to 2m_h$. The FFNS contributions,
$F_{2,h}^{\gamma}|_{\mathrm{dir}}$ (Eqs. (\ref{bhform}--\ref{bhformbis})) and
$F_{2,h}^{\gamma}|_{\mathrm{res}}$ (Eqs. (\ref{res}),(\ref{defchi})), behave 
properly at the thresholds. The problem emerges for the heavy-quark densities 
$q_h\gam$ and the subtraction terms (Eqs. (\ref{ss18}),(\ref{ss19})). These 
terms do not naturally disappear for $W \to 2m_h$. Fortunately, this 
problem can be cured; we noticed that the resolved-photon contribution in 
Eq. (\ref{res}) vanishes for $W\to 2m_h$ because then $\chi_h\to 1$ and 
the corresponding integral disappears. So, we can do the same with the
$q_h\gam$ distribution and the subtraction terms, if instead of $x$ we
introduce the $\chi_h$ variable (Eq. (\ref{defchi})) slightly shifted from $x$.
This way we force the heavy-quark distribution and the second term of the 
subtraction contribution (the integral term) to vanish at the corresponding 
threshold. Unfortunately, unlike for the proton, in the case of the photon we
are left with the $F_{2,h}^{\gamma}|_{\mathrm{dir},\mathrm{subtr}}$
contribution, which is now 
proportional to $\chi_h^2+(1-\chi_h)^2$ and does not vanish for $\chi_h \to 1$.
In the large-$Q^2$ region, where the ZVFNS is reliable, this change of 
variables is irrelevant. In the numerical calculations we ensure that the 
total contributions to $F_2^{\gamma}$ due to heavy quarks are not negative 
(positivity constraint). 
This way we effectively introduce small additional terms near the charm- and 
beauty-quark thresholds. The final formula for the \fun in the ACOT$(\chi)$ 
scheme reads\footnote{In our analysis we take the variable $\chi_h$ instead 
of $x$ in front of the bracket in the last sum on Eq. (\ref{finalacot}). The 
fit ($\chi^2$ value) agrees with the fit based on Eq. (\ref{finalacot}) at 
the per cent level. At the same time the value of $F_2^\gamma$ around 
thresholds is shifted by up to $\sim 10 \%$; also changing the form of the 
positivity constraint lead to changes of $F_2^\gamma$ in this region of 
similar size. Other subtraction terms and positivity constraints are under 
study and will be presented in a future publication}
\ba
F_2^{\gamma}(x,Q^2) &=& \sum_{i=1}^3 xe_i^2 (q_i\gam +\bar q_i\gam)(x,Q^2) +
\sum_{h(=c,b)}^2 xe_h^2 (q_h\gam + \bar q_h\gam)(\chi_h,Q^2) \nonumber \\
&& + \sum_{h(=c,b)}^2 \left[ F_{2,h}^{\gamma}(x,Q^2)|_{\mathrm{dir}} +
 F_{2,h}^{\gamma}(x,Q^2)|_{\mathrm{res}} \right] \\
&& - \sum_{h(=c,b)}^2 x \ln \frac{Q^2}{m_h^2} \left[e_h^4 
\frac{\alpha}{\pi}k(\chi_h) + e_h^2 \frac{\alpha_s(Q^2)}{\pi}\int_{\chi_h}^1 \frac{dy}{y} 
P_{qG} \left(\frac{\chi_h}{y}\right)G\gam(y,Q^2) \right]. \nonumber
\label{finalacot}
\ea

As can be seen the heavy-quark distributions are included in the second sum as
$q_h\gam(\chi_h,Q^2)$, being $\chi_h$ functions of $x$ (and $Q^2$).
We parametrize the final form obtained for these distributions in the Appendix 
as simple functions of $x$ and $Q^2$. \footnote{Which means that any further 
$x\to \chi_h$ substitution is not needed.}

The $\chi_h$ variables of the ACOT($\chi$) scheme recall the so-called 
``slow rescaling'' obtained in early papers on the charm-quark production in 
the DIS$_{ep}$ (e.g. \cite{slowresc}), where the Bjorken $x$ was replaced with
$\zeta$:
\be
x \to \zeta = x\left( 1+\frac{m_c^2}{Q^2} \right).
\end{equation}


\section{Global fits - solving the DGLAP evolution}

Using all the existing \fun data we perform two global fits, both based on 
the LO DGLAP evolution equations. One fit uses the ACOT$(\chi)$, the other a 
FFNS model for the heavy-quark contributions.

First we introduce the basic ingredients that are common for the two 
considered models.


\subsection{Mellin moments}

The LO DGLAP evolution equations are very much simplified if they are
transformed into the Mellin-moments space. The $n$-th moment for the quark or gluon 
densities, $f_i^{\gamma}$, is defined by
\begin{equation}
f_i^{\gamma,n}(Q^2) = \int_0^1 x^{n-1}f_i^{\gamma,n}(x,Q^2)dx.
\end{equation}
Analogous definitions can be used for the splitting functions $P_{ij}^n$. The 
evolution equations in the Mellin space have the generic form
\begin{equation}
\frac{d f_i^{\gamma,n}(Q^2)}{d\ln Q^2} =
\frac{\alpha}{2\pi}k_i^n(x) + \frac{\alpha_s(Q^2)}{2\pi}
P_{ij}^n f_j^{\gamma,n}(Q^2).
\label{apmel}
\end{equation}
Obviously, the first term on the right-hand side appears only for the quark
densities. For simplicity, in the following we will skip all subscripts and 
superscripts wherever possible.


\subsection{Non-singlet and singlet parton densities}

To solve the DGLAP equations we need to decompose the parton densities into 
the singlet and non-singlet (in flavour space) combinations. For the non-singlet
(ns) case we have
\be
\begin{split}
f_{\mathrm{ns}_{N_f}}^{\gamma}(Q^2) = \; &
\sum_{i=1}^{N_f} (e_i^2-\langle e^2 \rangle) \left[ q_i^{\gamma}(Q^2)
  + \bar{q}_i^{\gamma}(Q^2) \right],\\
k_{\mathrm{ns}_{N_f}} =\; &2N_f \left( \langle e^4 \rangle-{\langle
    e^2 \rangle}^2 \right) k
\label{nonsing}
\end{split}
\ee
where 
\be
\langle e^k \rangle \;=\; N_f^{-1}\sum_{i=1}^{N_f} e_i^k
\end{equation}
and $e_i$ stands for the corresponding quark electric charge. 
Similarly for the singlet (s) densities we have
\be
\begin{split}
f_\mathrm{s}^{\gamma}(Q^2) \;=\; &\left[ 
\begin{array}{c} \Sigma^{\gamma}(Q^2) \\ G^{\gamma}(Q^2) 
\end{array} \right], \\
\Sigma^{\gamma}(Q^2) \;=\; &\sum_{i=1}^{N_f}[q_i^{\gamma}(Q^2)+\bar
q_i^{\gamma}(Q^2)]. \qquad \qquad \!
\label{sing}
\end{split}
\end{equation}

In the singlet case the DGLAP equations (\ref{apmel}) become a matrix 
equation with
\be
\hat{P} = \left[ \begin{array}{cc} P_{qq} & 2 N_f P_{qG} \\
                                  P_{Gq} & P_{GG}
                       \end{array} \right], \qquad
\hat{k} = \left[ \begin{array}{c} k_\mathrm{s} \\ 0
                 \end{array} \right]
\end{equation}
with
\be
k_\mathrm{s} = 2N_f \langle e^2 \rangle k.
\end{equation}


\subsection{Point- and hadron-like parts}

The solution of the DGLAP equations can be divided into the so-called
point-like (pl) part, related to a special solution of the full inhomogeneous 
equation and hadron-like (had) part, arising as a general solution of the 
homogeneous equation. Their sum gives the partonic density in the photon, so 
we have
\be
f\gam(Q^2) = f_{\mathrm{had}}\gam(Q^2) + f_{\mathrm{pl}}\gam(Q^2)
\end{equation}
where
\ba
f_{\mathrm{pl}}\gam (Q^2) &=& \frac{4\pi}{\alpha_s(Q^2)}
\frac{1}{1-2P/\beta_0}\frac{\alpha}{2\pi \beta_0}
\left[1-L^{1-2P/\beta_0}\right] k, \nonumber \\
\quad & \quad & \quad \\
f_{\mathrm{had}}(Q^2) &=& L^{-2P/\beta_0}f\gam(Q_0^2). \nonumber
\label{ss31}
\ea
Here $\beta_0=11-2N_f/3$, $P$ equals $P_{qq}$ ($\hat{P}$) for the 
non-singlet (singlet) parton densities, and $L= \frac{\alpha_s(Q^2)}{\alpha_s(Q_0^2)}$, 
where the $Q_0^2$ is the evolution starting (input) scale. Note that at 
$Q^2=Q_0^2$ the point-like part vanishes.


\subsection{Input parton densities. VMD}

Following \cite{grv92}, the input scale has been chosen to be small,
$Q_0^2=0.25$ GeV$^2$, hence our parton densities are radiatively generated.
The point-like contributions are given by Eq. (30), while the hadronic 
parts need the input distributions. For this purpose we utilize the Vector 
Meson Dominance (VMD) model \cite{VMD}, where
\be
f_{\mathrm{had}}\gam(x,Q_0^2) = \sum_{V}\frac{4\pi \alpha}{\hat f^2_{V}}f^{V}(x,Q_0^2), 
\end{equation}
with the sum running over all light vector mesons (V) into which the photon 
can fluctuate. The parameters $\hat f^2_{V}$ can be extracted from the 
experimental data on the $\Gamma(V\to e^+e^-)$ width. In practice one takes 
into account the $\rho^0$ meson\footnote{An attempt to include the $J/\psi$ 
meson in the evolution has been made \cite{pjank}.} and the contributions
from other mesons are accounted for via a parameter $\kappa$, which is left 
as a free parameter. We take 
\begin{equation}
f_{\mathrm{had}}\gam(x,Q_0^2) = \kappa\frac{4\pi \alpha}{\hat f^2_{\rho}}f^{\rho}(x,Q_0^2).
\label{vmdfor}
\end{equation}

In the GRV prescription \cite{grv92} the parton densities in the
$\rho^0$ meson are approximated by the pionic ones: 
$f^{\rho} (x,Q^2_0) \approx f^{\pi} (x,Q^2_0)$. However, this assumption 
ignores among others the possible effects of the pseudo-Goldstone boson nature 
of the pion, and we are not using it in our analysis\footnote{Some
 indications that the valence distributions are indeed very
different in the $\pi$ and $\rho$ mesons
 have been recently given in \cite{joffe}.}. Instead we use the input 
densities of the $\rho^0$ meson at $Q_0^2=0.25$ GeV$^2$ in the form of 
valence-like distributions both for the (light) quark 
($v^{\rho}$) and gluon ($G^{\rho}$) densities.
All sea-quark distributions (denoted by $\zeta^{\rho}$) 
are neglected at the input scale. At this scale,
 the densities $v^{\rho}, G^{\rho}$ and $\zeta^{\rho}$ 
are related, according to Eq. (\ref{vmdfor}) to the corresponding
densities for a photon; see below.

The $v^{\rho}$ density is given by 
\be
v^{\rho}(x,Q_0^2)= \frac{1}{4}
(u^{\rho^+}+\bar u^{\rho^-}+d^{\rho^-}+\bar d^{\rho^+})(x,Q_0^2),
\end{equation}
where from the isospin symmetry
\be
u^{\rho^+}(x,Q_0^2) = \bar u^{\rho^-}(x,Q_0^2) 
= d^{\rho^-}(x,Q_0^2) = \bar d^{\rho^+}(x,Q_0^2).
\end{equation}
Note that all the densities in Eq. (35) are normalized to 1, 
\ie $\int^1_0 u^{\rho^+} dx =1$.

The following constraints should hold for the $v^{\rho}(x,Q_0^2)$ density.
The first is related to a number of valence quarks in the $\rho^0$ meson, 
and we have 
\be
\int_0^1 2 v^{\rho}(x,Q_0^2)dx = 2.
\label{const1}
\end{equation}
The second constraint represents the energy-momentum sum rule
\be
\int_0^1 x \left( 2v^{\rho}(x,Q_0^2)+G^{\rho}(x,Q_0^2) \right) dx = 1.
\label{const2}
\end{equation}

We parametrize the input densities as follows
\ba
x \zeta^{\rho}(x,Q_0^2)&=&0, \\
xv^{\rho}(x,Q_0^2) &=& N_v x^{\alpha}(1-x)^{\beta}, \\
xG^{\rho}(x,Q_0^2) &=& \tilde N_g xv^{\rho}(x,Q_0^2)= 
N_g x^{\alpha}(1-x)^{\beta},
\label{input1}
\ea
where $N_g=\tilde N_gN_v$, and impose two constraints
given by Eqs. (\ref{const1}) and (\ref{const2}) in both models. 
These constraints allow us to express the normalization factors $N_{v}$ and 
$N_g$ as functions of $\alpha, \beta$ and $\kappa$. This 
leaves these three parameters as the only free parameters to be fixed in the 
fits to the $F_2^{\gamma}$ experimental data.


\subsection{FFNS$_{CJKL}$ model}

In the FFNS$_{CJKL}$ model the number of ``active quarks'' ($N_f$) equals 3,
so $\langle e^2 \rangle =2/9$.

\bigskip

To describe the hadron-like part of the solution of DGLAP equations for the 
photon, we introduce, apart from the valence-like quark and gluon densities,
also the sea distribution $\zeta_{\mathrm{had}}^{\gamma}(x,Q^2)$. The up-, down- and 
strange-quark densities in the photon are then given by the following combinations

\ba
u_{\mathrm{had}}^{\gamma}(x,Q^2) &=& d_{\mathrm{had}}^{\gamma}(x,Q^2)
= \frac{1}{2} \left[
  v_{\mathrm{had}}^{\gamma}(x,Q^2)+2\zeta_{\mathrm{had}}^{\gamma}(x,Q^2) \right],
\label{grvcom1} \\
s_{\mathrm{had}}^{\gamma}(x,Q^2) &=& \zeta_{\mathrm{had}}^{\gamma}(x,Q^2).
\label{grvcom2}
\ea

>From Eqs. (26) and (\ref{sing}) we get (below we simply use
$f_{\mathrm{ns}}^{\gamma}$ instead of $f_{\mathrm{ns}_3}^{\gamma}$):
\ba
f_{\mathrm{ns},\mathrm{had}}^{\gamma}(x,Q^2) &=& \! \frac{1}{9}v_{\mathrm{had}}\gam(x,Q^2),\\
\Sigma_{\mathrm{had}}^{\gamma}(x,Q^2) &=& 2v_{\mathrm{had}}\gam(x,Q^2)
+ 6\zeta_{\mathrm{had}}\gam(x,Q^2).
\ea
After performing the DGLAP evolution of $f_{\mathrm{ns},\mathrm{had}}^{\gamma}(x,Q^2)$ and 
$\Sigma_{\mathrm{had}}^{\gamma}(x,Q^2)$ from $Q_0^2$ to higher $Q^2$, we calculate
$v_{\mathrm{had}}^{\gamma} (x,Q^2)$ and $\zeta_{\mathrm{had}}^{\gamma}(x,Q^2)$. Finally,
using formulae (\ref{grvcom1}) and (\ref{grvcom2}), we obtain the hadron-like part for the 
individual quark densities.

\bigskip

As the down- and strange-quarks have equal electric charges, there are only 
two different point-like distributions: $u_{\mathrm{pl}}\gam (x,Q^2)$ and 
$d_{\mathrm{pl}}\gam (x,Q^2)=s_{\mathrm{pl}}\gam(x,Q^2)$. We calculate them again through the 
evolution of the singlet and non-singlet combinations of the parton densities. 
It can be easily checked that distributions read as
\be
\begin{split}
u_{\mathrm{pl}}^{\gamma}(x,Q^2) = \frac{1}{\;6\;} \big[ \phantom{1}&\Sigma_{\mathrm{pl}}\gam(x,Q^2)+9f_{\mathrm{ns},\mathrm{pl}}\gam(x,Q^2) \big],\\
d_{\mathrm{pl}}^{\gamma}(x,Q^2) = \frac{1}{12} \big[ 2&\Sigma_{\mathrm{pl}}\gam(x,Q^2)-9f_{\mathrm{ns},\mathrm{pl}}\gam(x,Q^2) \big].
\end{split}
\end{equation}

\bigskip

Finally, the contribution due to the massive $c$- and $b$-quarks are 
approximated by the Bethe--Heitler formula (10)--(12) for $Q^2>4m_h^2$.


\subsection{CJKL model}

In the CJKL model all terms originating from both FFNS and ZVFNS are 
included. This means that apart from the light-quark distributions we take 
into consideration $c\gam(x,Q^2)$ and $b\gam(x,Q^2)$, which emerge in the 
ZVFNS, so here $N_f=5$. 

\bigskip
 
When five ``active quarks'' are considered instead of three, the DGLAP 
evolution becomes slightly more complicated and we need more
non-singlet parton densities than for the simple FFNS model. Here we need 
$f_{\mathrm{ns}_{2}},f_{\mathrm{ns}_{3}},f_{\mathrm{ns}_{4}}$, and
$f_{\mathrm{ns}_{5}}$ (Eq. (\ref{nonsing})) for both hadron- and
point-like parts; using them and $\Sigma$, calculated for $N_f=5$, we can 
express individual quark densities\footnote{We introduce below $c(x,Q^2)$ and
$b(x,Q^2)$ to describe densities of $c$- and $b$-quarks, respectively, like in 
any standard massless scheme with 5 flavours. Note also that formally, at the 
moment, we have all five densities at each $(Q^2,x)$. These ``initial'' 
densities for $c$- and $b$-quarks are later modified according to the ACOT 
prescription, leading to the final densities denoted by the same symbols; see 
main text for a detailed explanation.} as follows:
\be
\begin{split}
u\gam & = \frac{1}{20}(\phantom{+00f_{\mathrm{ns}_2}\gam} + 45f_{\mathrm{ns}_3}\gam - 30f_{\mathrm{ns}_4}\gam + 15f_{\mathrm{ns}_5}\gam + 2\Sigma\gam ),\\
d\gam & = \frac{1}{20}( - 60f_{\mathrm{ns}_2}\gam + 45f_{\mathrm{ns}_3}\gam - 30f_{\mathrm{ns}_4}\gam + 15f_{\mathrm{ns}_5}\gam + 2\Sigma\gam ),\\
s\gam & = \frac{1}{20}( + 60f_{\mathrm{ns}_2}\gam - 45f_{\mathrm{ns}_3}\gam - 30f_{\mathrm{ns}_4}\gam + 15f_{\mathrm{ns}_5}\gam + 2\Sigma\gam ),\\
c\gam & = \frac{1}{20}(\phantom{+00f_{\mathrm{ns}_2}\gam} - 45f_{\mathrm{ns}_3}\gam + 30f_{\mathrm{ns}_4}\gam + 15f_{\mathrm{ns}_5}\gam + 2\Sigma\gam ),\\
b\gam & = \frac{1}{10}(\phantom{+00f_{\mathrm{ns}_2}\gam -
 00f_{\mathrm{ns}_3}\gam} + 30f_{\mathrm{ns}_4}\gam - 30f_{\mathrm{ns}_5}\gam +
\phantom{1}\Sigma\gam ).
\end{split}
\label{udscb}
\end{equation}

For the hadron-like parts we consider, similarly to the FFNS case, the 
light-quarks densities given by Eq. (\ref{grvcom1}). Among sea quarks we have 
now all types of quarks, in particular we have:
\be
\begin{split}
s_{\mathrm{had}}^{\gamma}(x,Q^2) = c_{\mathrm{had}}^{\gamma}(x,Q^2) = 
b_{\mathrm{had}}^{\gamma}(x,Q^2)
&= \zeta_{\mathrm{had}}^{\gamma}(x,Q^2).
\end{split}
\end{equation}
This leads to the following relations
\be
\begin{split}
f_{\mathrm{ns}_2,\mathrm{had}}^{\gamma}=&\; f_{\mathrm{ns}_4,\mathrm{had}}^{\gamma}= 0 \\
f_{\mathrm{ns}_3,\mathrm{had}}^{\gamma}=&\; \frac{1}{\;9\;} v_{\mathrm{had}}^{\gamma},\\
f_{\mathrm{ns}_5,\mathrm{had}}^{\gamma}=&\; \frac{1}{15} v_{\mathrm{had}}^{\gamma}.
\end{split}
\ee
valid at every $x$ and $Q^2$.

In the point-like case equality of the electric charges for the up-type and 
down-type quarks leads to the following relations:
$u_{\mathrm{pl}}\gam(x,Q^2)=c_{\mathrm{pl}}\gam(x,Q^2)$ and $d_{\mathrm{pl}}\gam(x,Q^2)=
s_{\mathrm{pl}}\gam(x,Q^2)=b_{\mathrm{pl}}\gam(x,Q^2)$.

In our analysis, performed within the ACOT$(\chi)$ scheme, we evolve first the 
singlet and non-singlet distributions and we obtain in this way the gluon 
$g(x,Q^2)$ and light-quarks $u,d,s(x,Q^2)$ densities. In each heavy-quark 
density, \ie for $c$ and $b$ quarks, we replace $x$ variable by the 
corresponding $\chi_h$ variable; let us recall that 
$\chi_h = x\left(1+\frac{4m_h^2}{Q^2}\right)$, with $m_h$ equal to $m_c$ and $m_b$, respectively.
For such densities we perform DGLAP evolutions and obtain the resulting 
$c(\chi_h,Q^2)$ and $b(\chi_h,Q^2)$ distributions. We then compute the \fun 
using Eq. (\ref{finalacot}) and fit the parameters of the model to the experimental 
data, which is described in more detail in the next section.

\bigskip

Finally, we parametrize all our resulting parton distributions 
analytically with the possibly simple functions of $x$ and $Q^2$. 
This parametrization is given in the Appendix.


\section{Global fits and results}

We have performed two fits to all the available \fun data 
\cite{CELLO}--\cite{HQ2}. All together, 208 data points were used, including the 
recent high-$Q^2$ measurement of the OPAL collaboration \cite{HQ2}. Some of 
the data points are not in agreement with others. We will discuss in detail 
their influence on the fit in the next section. The fits based on the
least-squares principle (minimum of $\chi^2$) were done using \textsc{Minuit} \cite{minuit}. 
Systematic and statistical errors on data points were added in quadrature.

The $\alpha_s$ value used in the fits was calculated from the LO formula, which
depends on $N_f$
\be
\alpha_s(Q^2)^{(N_f)} = \frac{4\pi}{\beta_0 \ln (Q^2/\Lambda^{(N_f)})} \quad 
\mathrm{with} \quad \beta_0 = 11-\frac{2}{3}N_f.
\end{equation}
For $N_f=4$ we took the QCD scale $\Lambda^{(4)}$ equal to 280 MeV \cite{RPP}
with the assumption that the LO and NLO $\Lambda$ values for four active 
flavours are equal, which is consistent with the GRV group approach 
\cite{grv92}. 
Values of $\Lambda$ for other $N_f$ can be calculated if one assumes a 
continuity of $\alpha_s$ at the heavy-quark thresholds $Q^2=m_h^2$. Assuming 
then that $\alpha_s(m_h^2)^{(N_f)}=\alpha_s(m_h^2)^{(N_f+1)}$ one obtains the 
relation
\be
\Lambda^{(N_f+1)}=m_h(\Lambda^{(N_f)}/m_h)^{(33-2N_f)/(31-2N_f)},
\end{equation}
which gives $\Lambda^{(3)}=314$ MeV and $\Lambda^{(5)}=221$ MeV. Finally, the 
masses of the heavy quarks are taken to be \cite{RPP}: $m_c=1.3$ GeV and 
$m_b=4.3$ GeV.

The results of both fits are presented in Table \ref{tparam}. The second and third columns 
show the quality of the fits, i.e. the total $\chi^2$ for 208 points and the 
$\chi^2$ per degree of freedom. The fitted values for parameters $\alpha$, 
$\beta$ and $\kappa$ are presented in the middle of the table. In addition, 
the values for $N_v$ and $\tilde N_{g}$ obtained from these parameters using 
the constraints (\ref{const1}) and (\ref{const2}) are given in the last two 
columns.
\begin{table}[htb]
\begin{center}
\begin{tabular}{|c|@{} p{0.1cm} @{}|c|c|@{} p{0.1cm} @{}|c|c|c|@{} p{0.1cm} @{}|c|c|}
\hline
 Model    && $\chi^2$ & $\chi^2/_{\rm DOF}$ && $\kappa$ & $\alpha$ & $\beta$ && $N_v$ & $ \tilde{N_{g}}$ \\
\hline
\hline
 FFNS$_{CJKL}$ &&    471             & 2.30            && 1.726    & 0.465    & 0.127   && 0.504 & 1.384 \\
\hline
 CJKL          &&    430             & 2.10            && 1.097    & 0.876    & 2.403   && 2.644 & 2.882 \\
\hline
\end{tabular}
\caption{The $\chi^2$ for 208 points and parameters of the fits for FFNS$_{CJKL}$ and CJKL models.}
\label{tparam}
\end{center}
\end{table}

We see that the obtained $\chi^2$ per degree of freedom is better in the CJKL 
model than in the standard-type FFNS approach; however, it is not particularly 
good, owing the poor quality of some data used in the analyses. This fact has been 
already discussed in many papers, \eg \cite{klasen}, see also discussion in section 5.1.

The two fits to the same collection of data, although not very different as far
as $\chi^2$ is concerned, are obtained with very different sets of parameters. 
Note that $\kappa$ is close to 1 in the CJKL case, while for the other fit it
is closer to 2. If this parameter is close to 1, we have in practice only the 
$\rho$ contribution at the input scale. However this is not the whole story 
since the $\kappa N_v$ and $\kappa N_g$ give full normalizations of the 
valence-like quark and gluon densities in the $\gamma$. Now, the $N_v$ and 
$\tilde N_g$ are much smaller in the standard approach than in the CJKL model.
Finally, let us notice the large difference in both models, small and 
large $x$, of the fitted input densities, which correspond to very 
different $\alpha$ and $\beta$ parameters, respectively. The FFNS$_{CJKL}$ 
model has $\alpha$ close to the standard (Regge) one for a valence-quark 
density ($\alpha-1 \sim -0.5$); however its $\beta$, which governs the
large-$x$ behaviour, is very small, being far from 2, a standard prediction from the
quark-counting rule \cite{joffe}. On the other hand, the CJKL model gives 
$\beta$ closer to 2, while its $\alpha -1$ is close to zero.


\subsection{Comparison of the CJKL and FFNS$_{CJKL}$ fits with $F_2\gam$ data and other LO parametrizations}

The $\chi^2$ values, representing the quality of our LO fits, are compared in 
Table \ref{chi} with the corresponding $\chi^2$ obtained by us using the GRS 
LO \cite{grs} and SaS1D \cite{sas} parametrizations to describe the present 
$F_2^{\gamma}$ data. The comparison is performed for a set of 205 data 
points, i.e. excluding the points with $Q^2<0.26$ GeV$^2$ since they were not 
used in performing the GRS parametrization. The second column gives the number
of independent parameters in each model. The overall $\chi^2$ and $\chi^2/$DOF
values are given in the middle of the table for 205 data points. It is clear 
that both our fits give a better description of the experimental data than the
previous parametrizations. This could be expected since we are including more 
data in our fits. The CJKL model gives the lowest value of $\chi^2/$DOF, but 
it is still rather high. This may arise from the fact that we use all 
available data and, as it was stated (\eg \cite{klasen}), the data published 
by the TPC/{$2\gamma$} Collaboration \cite{TPC} are inconsistent with other 
measurements. We studied this point and in the last two columns of the table 
we present the $\chi^2$ values calculated without the TPC/{$2\gamma$} points. 
Indeed the $\chi^2/$DOF then computed is visibly improved\footnote{Note that
those data have been used in the fits.}. A special CJKL fit performed without 
TPC/{$2\gamma$} data gives $\chi^2$/DOF equal to 1.78. The very recent NLO 
analysis performed in \cite{klasen} for 134 experimental points and with five 
independent parameters gives $\chi^2/{\rm DOF} = 0.93$.

\begin{table}[h]
\begin{center}
\begin{tabular}{|c|@{} p{0.1cm} @{}|c|@{} p{0.1cm} @{}|c|c|@{} p{0.1cm} @{}|c|c|}
\hline
\hline
               &&                 && \multicolumn{5}{c|}{\# of data points} \\
\cline{5-9}
model          && \# of ind. par. && \multicolumn{2}{c|}{205} && 
\multicolumn{2}{c|}{182 - no TPC} \\
\cline{5-9}
               &&                 && $\chi^2$ & $\chi^2/_{\rm DOF}$ && $\chi^2$ & $\chi^2/_{\rm DOF}$ \\

\hline
\hline
SaS1D          &&    6            && 657                & 3.30            && 611& 3.47 \\
\hline
GRS LO         &&    0            && 499                & 2.43            && 366& 2.01\\
\hline
FFNS$_{CJKL}$  &&    3            && 442                & 2.19            && 357& 1.99\\
\hline
CJKL           &&    3            && 406                & 2.01            && 323& 1.80\\
\hline
\hline
\end{tabular}
\end{center}
\caption{Comparison of $\chi^2$ values obtained for the FFNS$_{CJKL}$ and CJKL 
fits to the \fun data with $\chi^2$ calculated using the SaS1D and GRS
parametrizations.}
\label{chi}
\end{table}

Figures \ref{fit1}--\ref{fit4} show a comparison of the CJKL and FFNS$_{CJKL}$ 
fits to \fun with the experimental data as a function of $x$, for
different values of $Q^2$. Also a comparison with the GRS LO and SaS1D parametrizations 
is shown. (If a few values of $Q^2$ are displayed in a panel, the average of 
the smallest and biggest one was taken in the computation.) As can be seen 
in Figs. \ref{fit1} and \ref{fit2}, both CJKL and FFNS$_{CJKL}$ models 
predict a much steeper behaviour of the \fun at small $x$ with respect to other 
parametrizations. In the region of $x \gtrsim 0.1$, the behaviour of the \fun 
obtained from the FFNS$_{CJKL}$ fit is similar to the ones predicted by the 
GRS LO and SaS1D parametrizations. The CJKL model gives lower prediction 
whenever the charm-quark threshold is surpassed, and slightly below this 
region, as is shown in Figs. \ref{fit3} and \ref{fit4}.

Apart from this direct comparison with the photon structure-function data,
we perform another comparison, this time with LEP data that were not used 
directly in our analysis. Figures \ref{evol1} and \ref{evol2} present the 
predictions for \fund, averaged over various $x$ regions, compared with the 
recent OPAL data \cite{HQ2}. For comparison, the results from the GRS LO and 
SaS1D parametrizations are shown as well. We observe that all models give very 
similar predictions, which are in fairly good agreement with the experimental 
data. Only the CJKL model slightly differs from the other models
considered: it gives better agreement with the data. The difference between the 
predictions of the CJKL model and the other models is most striking in the 
case of the medium-$x$ range, $0.1<x<0.6$ shown in Fig. \ref{evol1}. The 
CJKL curve clearly shows a departure from the simple $\ln Q^2$ dependence. 
This is caused by the additional $Q^2$ dependence due to the $\chi$ variable.
The highest $x$ range, $0.85<x<0.98$ (see Fig. \ref{evol2}) is the second 
region of a significant difference between the predictions of the CJKL and 
other models. The predictions split at high-$Q^2$ values, as the CJKL predicts
a much softer $Q^2$ dependence.


\subsection{Parton densities}

In this section we present the parton densities obtained from the CJKL and 
FFNS$_{CJKL}$ fits and compare them with the corresponding distributions of 
the GRV LO \cite{grv92}, GRS LO and SaS1D parametrizations. First, we present 
all parton densities at $Q^2=10$ GeV$^2$ (Fig. \ref{parton}). The biggest 
difference between our CJKL model and others is observed, as 
expected, for the heavy-quark distributions. Unlike for the GRV LO and SaS1D 
parametrizations, the densities $c^{\gamma}(x,Q^2)$ and $b^{\gamma}(x,Q^2)$ 
vanish not at $x=1$ but, as it should be, at the kinematical threshold. Also 
the up-quark density differs among models. In the CJKL model it is lower than 
in other parametrizations for $x>0.1$. The same can be seen in Fig. \ref{up}, 
where for various $Q^2$ values the up-quark distributions are presented. The 
up-quark density in the CJKL, FFNS$_{CJKL}$ and GRV LO models have similar 
behaviour at very small $x$. The hardest up-quark distribution is obtained in 
the FFNS$_{CJKL}$ approach, while both the GRS LO and SaS1D predictions are 
much softer. The same holds in the case of the gluon distribution, shown in 
Fig. \ref{glu}. Finally, the charm-quark densities of the CJKL model and of 
the GRV LO and GRS LO parametrizations are presented in Fig. \ref{chm}. Here 
we see that, in addition to the already mentioned vanishing at the threshold, 
the charm-quark distribution in the CJKL model is larger than the ones in the 
other parametrizations. This is particularly true for larger values of $Q^2$, 
where the threshold is very close to $x=1$.

Finally, in Fig. \ref{acot} we present our predictions for the $F_{2,c}\gam$. 
For $Q^2=5,20,100$ and {1000} GeV$^2$ we compare the individual contributions
included in the CJKL model. As was explained in detail in section 3.2, 
they all, apart from the $F_{2,c}|_{\mathrm{subtr},\mathrm{dir}}$
term, vanish in the $W\to 2m_c$ threshold. This term dominates near
the highest kinematically allowed $x$. The direct BH term is important
in the medium-$x$ range. Its shape resembles the valence-type
distribution. The charm-quark density contribution, \ie the term 
$2xe_c^2 c^{\gamma}(x,Q^2)$, is important in the whole kinematically available 
$x$ range; it dominates the $F_{2,c}\gam$ for small $x$. In this region also 
the resolved-photon contributions increase, but they cancel each other.


\subsection{Comparison with $F\gam_{2,c}$}

A good test of the charm-quark contributions is provided by the OPAL 
measurement of the $F_{2,c}\gam$, obtained from the inclusive production of 
$D^{*\pm}$ mesons in photon--photon collisions \cite{F2c}. The averaged 
$F_{2,c}\gam$ has been determined in the two $x$ bins. These data points are 
compared to the predictions of the CJKL and FFNS$_{CJKL}$ models as
well as with the SaS1D and GRS LO parametrizations in Fig. \ref{fF2c}. The prediction
of the FFNS$_{CJKL}$ model, which as the only one among those compared
does not contain the resolved-photon contribution, is based on the point-like 
(Bethe--Heitler) contribution for heavy quarks only. As seen in the figure it 
decreases too quickly with the decreasing $x$, much faster than the 
predictions from other parametrizations. The CJKL model seems to give the best
description of the data for the low-$x$ bin, but overshoots the experimental 
point at high $x$.


\section{Summary and outlook}

A new analysis of the radiatively generated parton distributions in the real 
photon based on the LO DGLAP equations is presented. All available 
experimental data have been used to perform two global, 3-parameter fits. 
Our main model (CJKL) is based on a new variable flavour-number scheme 
(ACOT(${\chi}$)), applied to the photon case for the very first time. It has 
a proper threshold behaviour of the heavy-quark contributions. The CJKL-model 
results are compared with an updated Fixed Flavour-Number Scheme (FFNS$_{CJKL}$)
fit and to predictions of the GRS LO and SaS1D parametrizations. Our model 
gives the best $\chi^2$ of those compared. It describes very well the $Q^2$
evolution of the \fun, averaged over various $x$-regions. We have checked that 
the CJKL fit agrees also reasonably well with the prediction of a sum rule for
the photon, described in \cite{sumrule}.

We have also checked that the gluon densities of both CJKL and FFNS$_{CJKL}$ 
models agree with the H1 measurement of the gluon density ($G^{\gamma}$)
performed at $Q^2=74$ GeV$^2$ \cite{h1glu}. In both models gluon densities are
very similar to the gluon density provided by the GRV LO parametrization, 
which gave so far the best agreement with the H1 data. Further comparison of 
our gluon densities to the H1 data cannot be performed in a fully consistent
way\footnote{A plot illustrating the comparison can be found in \cite{pjank1}.}
, since GRV LO proton and photon parametrization were used in the experiment
in order to extract such gluon density.

One of the motivations for this work was given by the disagreement between the
theoretical and experimental results for the open beauty-quark production in 
two-photon processes in the LEP \cite{bqOPAL,bqL3} measurements. We did 
calculate the LO cross-section for charm- and beauty-quark production in 
$\gamma \gamma$ collisions in the ACOT($\chi$) and FFNS schemes, using the 
CJKL and FFNS$_{CJKL}$ distributions of partons, respectively. The 
cross-section for the $c$-quark production computed in both models agrees with 
the experimental data. The ACOT($\chi$) model gives a slightly better shape of 
the $c$-quark distribution. There is a small difference between the results of
the two models for $b$-quark production. We observe an increase of the 
cross-section for the beauty-quark in the ACOT($\chi$) approach, as compared 
to the FFNS result, based on GRS LO parametrization, but it is too small to fit
the experimental data. More work on this subject is required. Also, before 
reaching a definitive conclusion the NLO corrections should be considered and 
the NLO parton densities for the photon applied. The NLO parametrization in 
the CJKL model, together with the effects of different subtraction terms and 
positivity constraints, will be presented in a future publication.

A simple analytic parametrization of the results of our CJKL model for the 
individual parton densities is given, and a \textsc{fortran} program calculating parton
densities as well as the structure function $F_2^{\gamma}$ (Eq. (\ref{finalacot})) can be 
obtained from the web sites given in \cite{webprog}.


\begin{acknowledgments}

MK thanks Wu-ki Tung for fruitful discussions, which led to this analysis, and 
R. Roberts for a discussion on the parton densities in the $\rho$-meson.
PJ would like to thank R.Taylor, M.Wing and all members of the Warsaw TESLA 
group for discussions. We are grateful to Rohini Godbole for useful suggestions 
on the parton-parametrization program and Mariusz Przybycie\'n for discussions.
This work was partly supported by the European Community's Human Potential 
Programme under contract HPRN-CT-2000-00149 Physics at Collider
and HPRN-CT-2002-00311 EURIDICE. FC also acknowledges partial financial
support by MCYT under contract FPA2000-1558 and Junta de Andaluc{\'\i}a group 
FQM 101. MK is grateful for partial support by Polish Committee for Research 
2P03B05119 and 5P03B12120.

\end{acknowledgments}


\appendix*
\section{Parton parametrization for the CJKL model}

We give here an analytic form for the parametrization of the CJKL results for
the individual parton densities. Following the GRV group we 
parametrize them in terms of
\be
s \equiv \ln \frac{\ln[Q^2/(0.221\; \mathrm{GeV})^2]}
                  {\ln[Q_0^2/(0.221\; \mathrm{GeV})^2]}
\end{equation}
with $Q_0^2=0.25$ GeV$^2$. The parametrization has been performed in the 
$10^{-5}<x<1$ and $1<Q^2<2\times 10^5$ GeV$^2$ ranges. We made a separate 
parametrization of the point- and hadron-like densities. The parametrized 
distributions of the light ($u,d,s$) quarks are in agreement with the ones 
obtained in the fit up to few percent accuracy. The same is true for the 
gluon density, apart from the high-$x$ region (where $G_{\gamma}$ values fall 
down), where 10\% accuracy is assured. We checked that the heavy-quark 
densities, for $c$- and $b$-quarks, are represented by our parametrization at 10\% accuracy for 
$Q^2\ge 2$ GeV$^2$ and $10$ GeV$^2$ respectively. For both quarks
those $Q^2$ values are below their masses squared 
$m_h^2=1.3^2 ~{\rm and } ~4.3^2$ GeV$^2$, which are often considered as the 
energy scale of the processes involving heavy quarks. The $\chi^2$ obtained 
for 205 data points for \fun is equal 406, for both fitted and 
parametrized distributions.

The final parton densities in the real photon are (we skip the superscript
$\gamma$ in this part):
\ba
G(x,Q^2) &=& G_{\mathrm{pl}}(x,Q^2) + G_{\mathrm{had}}(x,Q^2) \nonumber \\
d(x,Q^2) &=& d_{\mathrm{pl}}(x,Q^2) + \frac{1}{2} v(x,Q^2) + \zeta(x,Q^2) \nonumber \\
u(x,Q^2) &=& u_{\mathrm{pl}}(x,Q^2) + \frac{1}{2} v(x,Q^2) + \zeta(x,Q^2) \nonumber \\
s(x,Q^2) &=& s_{\mathrm{pl}}(x,Q^2) + \zeta(x,Q^2) \\
c(x,Q^2) &=& c_{\mathrm{pl}}(x,Q^2) + c_{\mathrm{had}}(x,Q^2) \nonumber \\
b(x,Q^2) &=& b_{\mathrm{pl}}(x,Q^2) + b_{\mathrm{had}}(x,Q^2) \nonumber
\ea

Note that all our densities describe the massless partons, although the 
kinematical constraints for $c$- and $b$-quarks were taken into account. The 
formulae given in the Appendix parametrizing densities of both heavy quarks, 
represent the densities as included in the second sum in the Eq. (\ref{finalacot}), that 
means that, for instance, the final density $b(x,Q^2)$ should be understood as 
being equivalent to $b(\chi_h(x,Q^2),Q^2)$.

A \textsc{fortran} code of the parametrization can be obtained from the web pages
\cite{webprog}. The program includes also an option for the \fun calculation
according to the Eq. (\ref{finalacot}).



\subsection{Point-like part}

We use the following form for the parametrization of the point-like distribution
for light quarks and gluons (denoted by $f_{\mathrm{pl}}(x,Q^2)$)

\ba
\frac{1}{\alpha}xf_{\mathrm{pl}}(x,Q^2) &=& \frac{9}{4\pi}\ln\frac{Q^2}{(0.221\; 
\mathrm{GeV})^2} \times \\ 
&& \left[s^{\alpha}x^a(A+B\sqrt{x}+Cx^b)
+ s^{\alpha'}\exp\left(-E + \sqrt{E' s^{\beta}\ln\frac{1}{x}}\right)\right]
(1-x)^D. \nonumber
\ea

For the gluon density $G_{\mathrm{pl}}$:

\be
\begin{array}{l}
\begin{array}{lllllllllll}
\alpha &=& -0.43865 &,& \alpha' &=& 2.7174 &,& \beta &=& 0.36752, \\
\end{array}\\
\begin{array}{l}
A = 0.086893  - 0.34992 s , \\
B = 0.010556 + 0.049525 s , \\
C = -0.099005 + 0.34830 s , \; D = 1.0648   + 0.143421 s, \\
E = 3.6717    + 2.5071  s , \; E'= 2.1944   + 1.9358   s, \\
a = 0.23679   - 0.11849 s , \; b = -0.19994 + 0.028124 s.
\end{array}
\end{array}
\end{equation}\\
For the up-quark density $u_{\mathrm{pl}}$:

\be
\begin{array}{l}
\begin{array}{lllllllllll}
\alpha &=& -1.0711 &,& \alpha' &=& 3.1320 &,& \beta &=& 0.69243,
\end{array}\\
\begin{array}{l}
A = -0.058266  + 0.20506  s , \\
B  = 0.0097377 - 0.10617  s , \\
C = -0.0068345 + 0.15211  s , \\
D  = 0.22297   + 0.013567 s , \\
E = 6.4289     + 2.2802   s , \; E' = 1.7302 + 0.76997  s, \\
a = 0.87940    - 0.110241 s , \; b  = 2.6878 - 0.040252 s.
\end{array}
\end{array}
\end{equation}\\
For the down- and strange-quark densities, $d_{\mathrm{pl}}=s_{\mathrm{pl}}$:

\be
\begin{array}{l}
\begin{array}{lllllllllll}
\alpha &=& -1.1357 &,& \alpha' &=& 3.1187 &,& \beta &=& 0.66290,
\end{array}\\
\begin{array}{l}
A = 0.098814   - 0.067300 s ,\\
B = -0.092892 + 0.049949 s  ,\\
C = -0.0066140 + 0.020427 s ,\\
D = -0.31385  - 0.0037558 s ,\\
E = 6.4671     + 2.2834   s , \; E' = 1.6996    + 0.84262   s, \\
a = 11.777     + 0.034760 s , \; b  = -11.124   - 0.20135   s.
\end{array}
\end{array}
\end{equation}\\

In the case of heavy quarks a slightly modified function $h_{\mathrm{pl}}$ was applied:

\ba
\frac{1}{\alpha}xh_{\mathrm{pl}}(x,Q^2) &=& \frac{9}{4\pi}\ln\frac{Q^2}{(0.221\; 
\mathrm{GeV})^2}\times \\
&& \left[s^{\alpha}y^a(A+B\sqrt{y}+Cy^b) + s^{\alpha'}\exp\left(-E + \sqrt{E' s^{\beta}\ln\frac{1}{x}}\right)\right](1-y)^D, \nonumber
\ea
with $y = x + 1 - \frac{Q^2}{Q^2+6.76 \mathrm{GeV}^2}$ for the charm-quark and
$y = x + 1 - \frac{Q^2}{Q^2+73.96 \mathrm{GeV}^2}$ for the beauty-quark densities.

\bigskip

For the charm-quark density $c_{\mathrm{pl}}$, for $Q^2 \le 10$ GeV$^2$:

\be
\begin{array}{l}
\begin{array}{lllllllllll}
\alpha &=& 2.9808 &,& \alpha' &=& 28.682 &,& \beta &=& 2.4863,
\end{array}\\
\begin{array}{l}
A = -0.18826   + 0.13565  s, \; B = 0.18508  - 0.11764 s, \\
C = -0.0014153 - 0.011510 s, \\
D = -0.48961 + 0.18810 s, \\
E = 0.20911 - 2.8544 s + 14.256 s^2, \\
E' = 2.7644 + 0.93717 s, \; a = -7.6307 + 5.6807 s, \\
b = 394.58 - 541.82 s + 200.82 s^2.
\end{array}
\end{array}
\end{equation}\\
For the charm-quark density $c_{\mathrm{pl}}$, for $Q^2 > 10$ GeV$^2$:

\be
\begin{array}{l}
\begin{array}{lllllllllll}
\alpha &=& -1.8095 &,& \alpha' &=& 7.9399 &,& \beta &=& 0.041563,
\end{array}\\
\begin{array}{l}
A = -0.54831 + 0.33412 s, \; B  = 0.19484 + 0.041562 s, \\
C = -0.39046 + 0.37194 s, \; D  = 0.12717 + 0.059280 s, \\
E =  8.7191  + 3.0194  s, \; E' = 4.2616  + 0.73993  s, \\
a = -0.30307 + 0.29430 s, \; b  = 7.2383  - 1.5995   s.
\end{array}
\end{array}
\end{equation}\\
For the beauty-quark density $b_{\mathrm{pl}}$, for $Q^2 \le 100$ GeV$^2$:

\be
\begin{array}{l}
\begin{array}{lllllllllll}
\alpha &=& 2.2849 &,& \alpha' &=& 6.0408 &,& \beta &=& -0.11577,
\end{array}\\
\begin{array}{l}
A = -0.26971 + 0.17942 s, \\
B = 0.27033 - 0.18358 s + 0.0061059 s^2, \\
C = 0.0022862 - 0.0016837 s, \\
D = 0.30807 - 0.10490 s, \; E = 14.812 - 1.2977 s, \\
E' = 1.7148 + 2.3532 s + 0.053734 \sqrt{s}, \\
a = 3.8140 - 1.0514 s, \; b = 2.2292 + 20.194 s
\end{array}
\end{array}
\end{equation}\\
For the beauty-quark density $b_{\mathrm{pl}}$, for $Q^2 > 100$ GeV$^2$:

\be
\begin{array}{l}
\begin{array}{lllllllllll}
\alpha &=& -5.0607 &,& \alpha' &=& 16.590 &,& \beta &=& 0.87190,
\end{array}\\
\begin{array}{l}
A = -0.72790  + 0.36549  s, \; B  = -0.62903 + 0.56817 s, \\
C = -2.4467   + 1.6783   s, \; D  = 0.56575  - 0.19120 s, \\
E =  1.4687   + 9.6071   s, \; E' = 1.1706   + 0.99674 s, \\
a = -0.084651 - 0.083206 s, \; b  =  9.6036  - 3.4864  s.
\end{array}
\end{array}
\end{equation}


\subsection{Hadron-like part}

We use a simple formula for the valence-quark density:

\be
\frac{1}{\alpha}xv(x,Q^2) = Ax^a(1+B\sqrt{x}+Cx)(1-x)^D,
\end{equation}
with the following parameters:
\be
\begin{array}{l}
\begin{array}{l}
A = 1.0898  + 0.38087  s, \; B = 0.42654 - 1.2128 s, \\
C = -1.6576 + 1.7075   s, \; D = 0.96155 + 1.8441 s, \\
a = 0.78391 - 0.068720 s.
\end{array}
\end{array}
\end{equation}\\

For the gluon distribution, we apply

\ba
\frac{1}{\alpha}xG_{\mathrm{had}}(x,Q^2) &=& \left[ x^a(A+B\sqrt{x}+Cx) \right. \\
&+& s^{\alpha}\exp\left(-E + \sqrt{E' s^{\beta}\ln\frac{1}{x}}\right)
\left. \right](1-x)^D, \nonumber
\ea

with

\be
\begin{array}{l}
\begin{array}{lllllll}
\alpha &=& 0.59945 &,& \beta &=& 1.1285, 
\end{array}\\
\begin{array}{l}
A = -0.19898 + 0.57414 s, \; B  = 1.9942  - 1.8306  s, \\
C = -1.9848  + 1.4136  s, \; D  = 0.21294 + 2.7450  s, \\
E =  1.2287  + 2.4447  s, \; E' = 4.9230  + 0.18526 s, \\
a = -0.34948 + 0.47058 s, \; b  = 1.0012  + 0.99767 s.
\end{array}
\end{array}
\end{equation}\\

In the case of the sea-quark density, we use:

\ba
\frac{1}{\alpha}x\zeta(x,Q^2) &=&  
\frac{s^{\alpha}}{\ln^a\frac{1}{x}} (1+A\sqrt{x}+Bx) \times \\
&& \exp\left(-E + \sqrt{E' s^{\beta}\ln\frac{1}{x}}\right)
(1-x)^D, \nonumber
\ea
with
\be
\begin{array}{l}
\begin{array}{lllllll}
\alpha &=& 0.71660 &,& \beta &=& 1.0497,
\end{array}\\
\begin{array}{l}
A  = 0.60478 + 0.036160 s, \; B = 4.2106  - 0.85835 s, \\
D  = 4.1494  + 0.34866  s, \; E = 4.5179  + 1.9219  s, \\
E' = 5.2812  - 0.15200  s, \; a = 0.72289 - 0.21562 s.
\end{array}
\end{array}
\end{equation}\\

Finally, for the heavy-quark densities:

\ba
\frac{1}{\alpha}xh_{\mathrm{had}}(x,Q^2) &=&  
\frac{s^{\alpha}}{\ln^a\frac{1}{x}} (1+A\sqrt{y}+By) \times \\
&& \exp\left(-E + \sqrt{E' s^{\beta}\ln\frac{1}{x}}\right) 
(1-y)^D, \nonumber
\ea

with $y = x + 1 - \frac{Q^2}{Q^2+6.76 \mathrm{GeV}^2}$ for charm- and
$y = x + 1 - \frac{Q^2}{Q^2+73.96 \mathrm{GeV}^2}$ for beauty-quark.

\bigskip

For the charm-quark density $c_{\mathrm{had}}$, for $Q^2 \le 10$ GeV$^2$:

\be
\begin{array}{l}
\begin{array}{lllllll}
\alpha &=& 5.6729 &,& \beta &=& 1.4575,
\end{array}\\
\begin{array}{l}
A  = -2586.4 + 1910.1 s, \; B = 2695.0  - 1688.2  s, \\
D  = 1.5146  + 3.1028 s, \; E = -3.9185 + 11.738  s,  \\
E' = 3.6126  - 1.0291 s, \; a = 1.6248  - 0.70433 s.
\end{array}
\end{array}
\end{equation}\\
For the charm-quark density $c_{\mathrm{had}}$, for $Q^2 > 10$ GeV$^2$:

\be
\begin{array}{l}
\begin{array}{lllllll}
\alpha &=& -1.6470 &,& \beta &=& 0.72738,
\end{array}\\
\begin{array}{l}
A = -2.0561 + 0.75576 s, \; B = 2.1266  + 0.66383 s, \\ 
D = 3.0301 - 1.7499 s + 1.6466  s^2 \\
E = 4.1282 + 1.6929 s - 0.26292 s^2,\\
E' = 0.89599 + 1.2761 s -0.15061 s^2, \\
a = -0.78809 + 0.90278 s.
\end{array}
\end{array}
\end{equation}\\
For the beauty-quark density $b_{\mathrm{had}}$, for $Q^2 \le 100$ GeV$^2$:

\be
\begin{array}{l}
\begin{array}{lllllll}
\alpha &=& -10.210 &,& \beta &=& -2.2296,
\end{array}\\
\begin{array}{l}
A  = -99.613 + 171.25   s, \; B = 492.61  - 420.45   s, \\
D  = 3.3917  + 0.084256 s, \; E = 5.6829  - 0.23571  s, \\
E' = -2.0137 + 4.6955   s, \; a = 0.82278 + 0.081818 s.
\end{array}
\end{array}
\end{equation}\\
For the beauty-quark density $b_{\mathrm{had}}$, for $Q^2 > 100$ GeV$^2$:

\be
\begin{array}{l}
\begin{array}{lllllll}
\alpha &=& 2.4198 &,& \beta &=& 0.40703,
\end{array}\\
\begin{array}{l}
A = -2.1109 + 1.2711 s, \; B = 9.0196 - 3.6082 s, \\
D = 3.6455  - 4.1353 s + 2.3615 s^2 , \\
E = 4.6196 + 2.4212 s, \; E' = 0.66454 + 1.1109 s, \\
a = -0.98933 + 0.42366 s + 0.15817 s^2. 
\end{array}
\end{array}
\end{equation}




\clearpage

\begin{figure}
\includegraphics[scale=1.0]{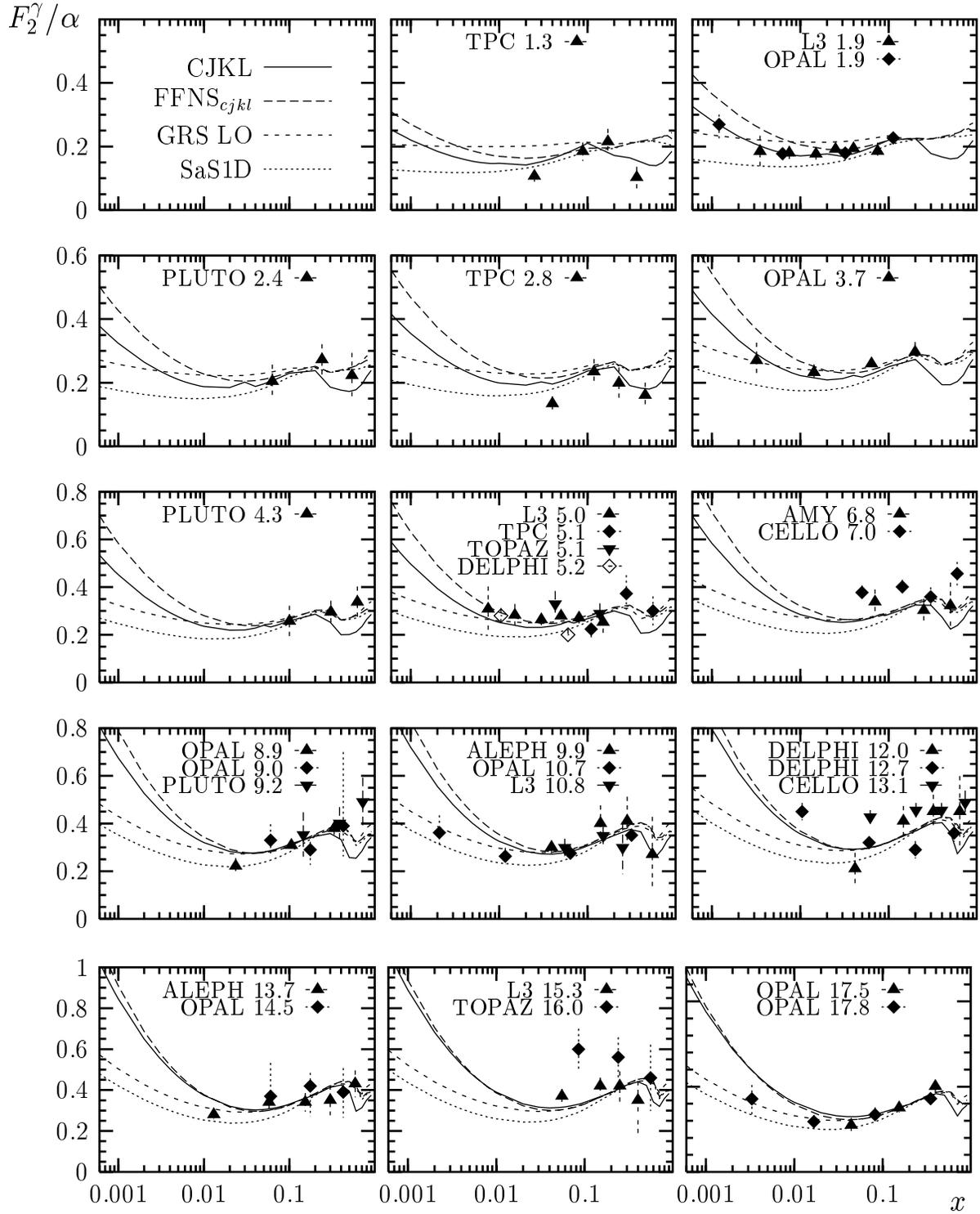}%
\caption{Predictions for the $F_2^{\gamma}(x,Q^2)/\alpha$ for the CJKL 
and FFNS$_{cjkl}$ models and GRS LO \cite{grs} and SaS1D \cite{sas} 
parametrizations compared with the experimental data \cite{CELLO}--\cite{HQ2},
for small and medium $Q^2$ as a function of $x$ (logarithmic scale). If a 
few values of $Q^2$ are displayed in the panel, the average of the smallest 
and biggest $Q^2$ was taken in the computation. }
\label{fit1}
\end{figure}

\newpage

\begin{figure}
\includegraphics[scale=1.0]{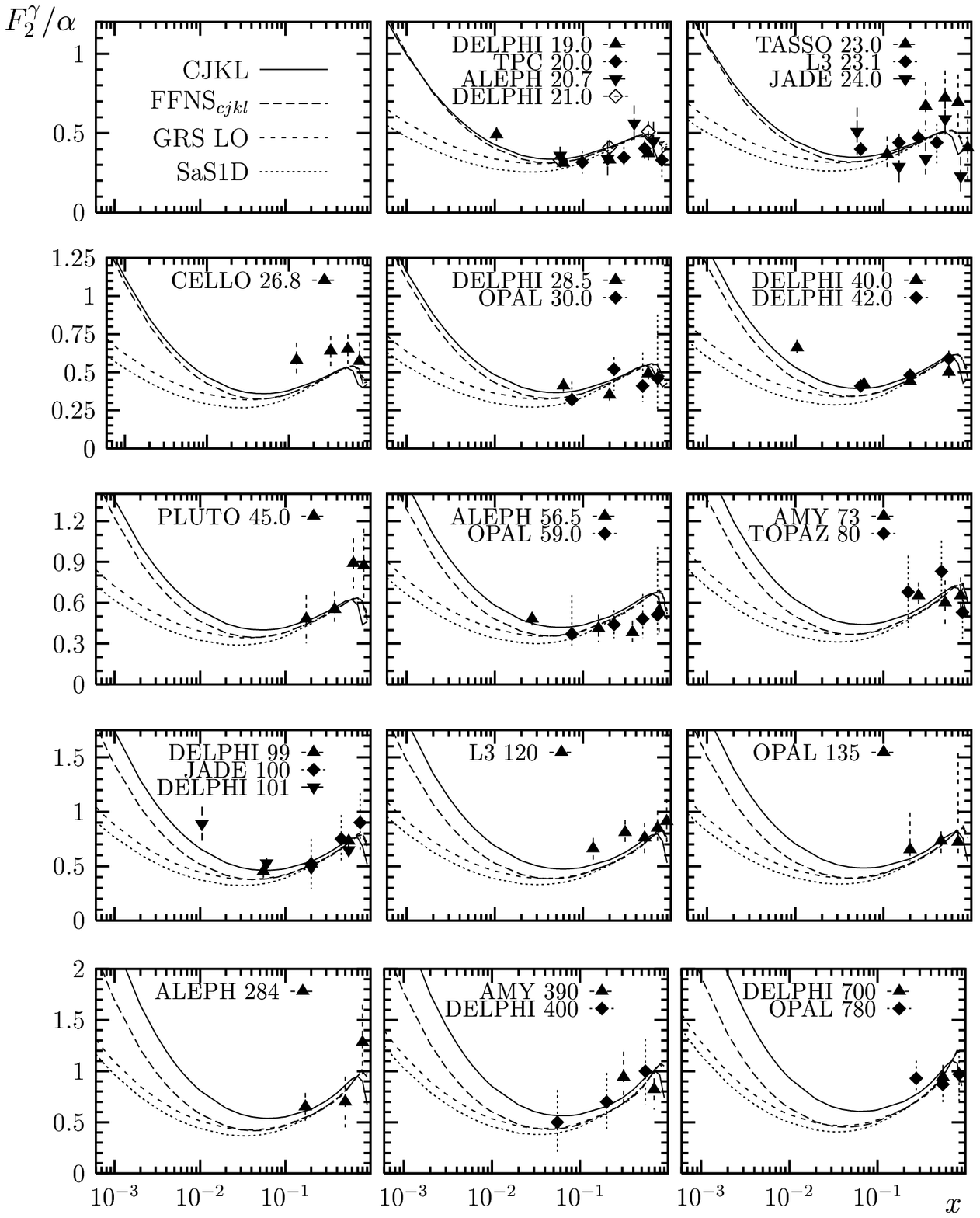}%
\caption{The same as in Fig. \ref{fit1}, for $Q^2 \gtrsim 20 \mathrm{GeV}^2$.}
\label{fit2}
\end{figure}

\newpage

\begin{figure}
\includegraphics[scale=1.0]{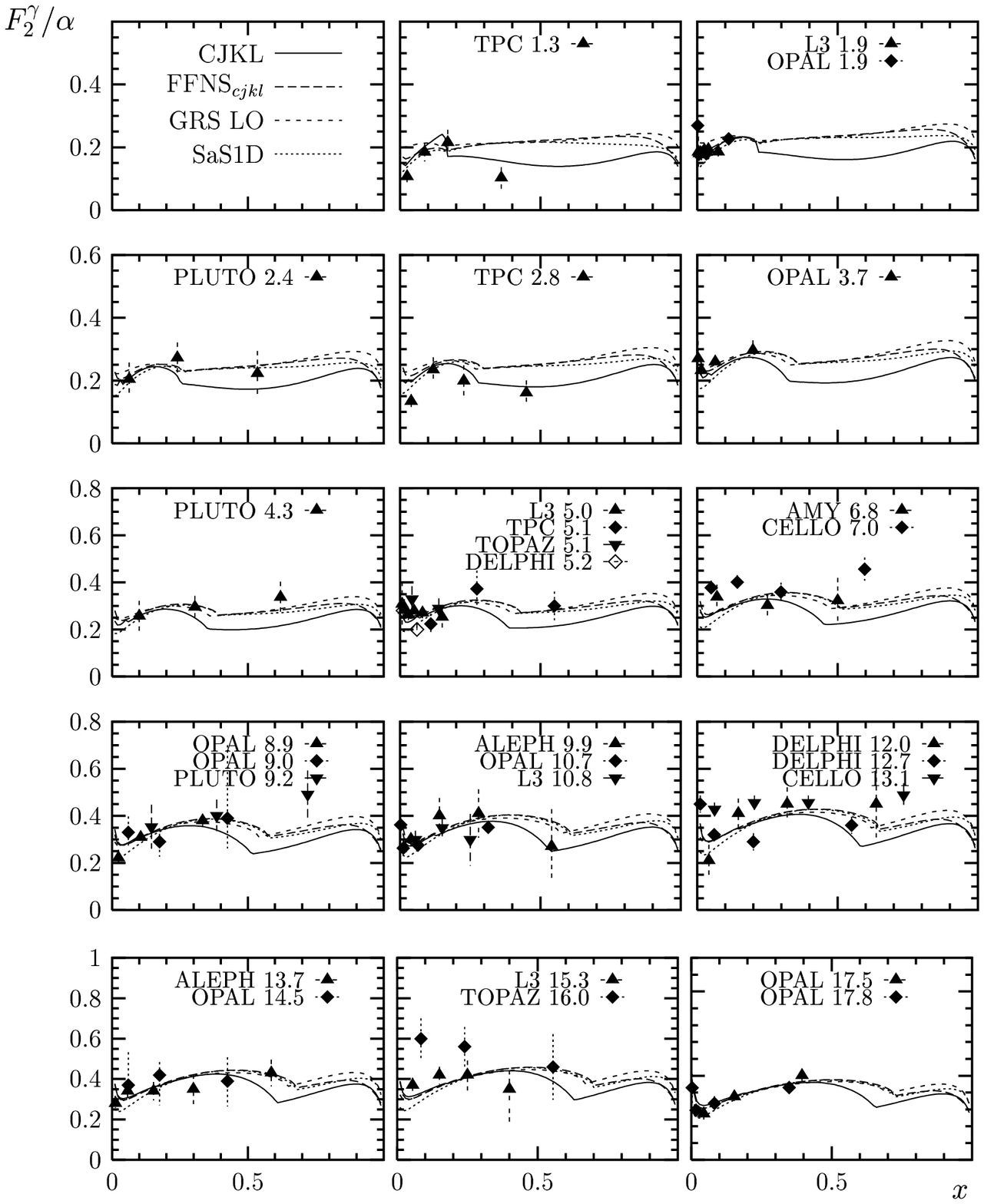}%
\caption{The same as in Fig. \ref{fit1} for a linear scale in $x$.}
\label{fit3}
\end{figure}

\newpage

\begin{figure}
\includegraphics[scale=1.0]{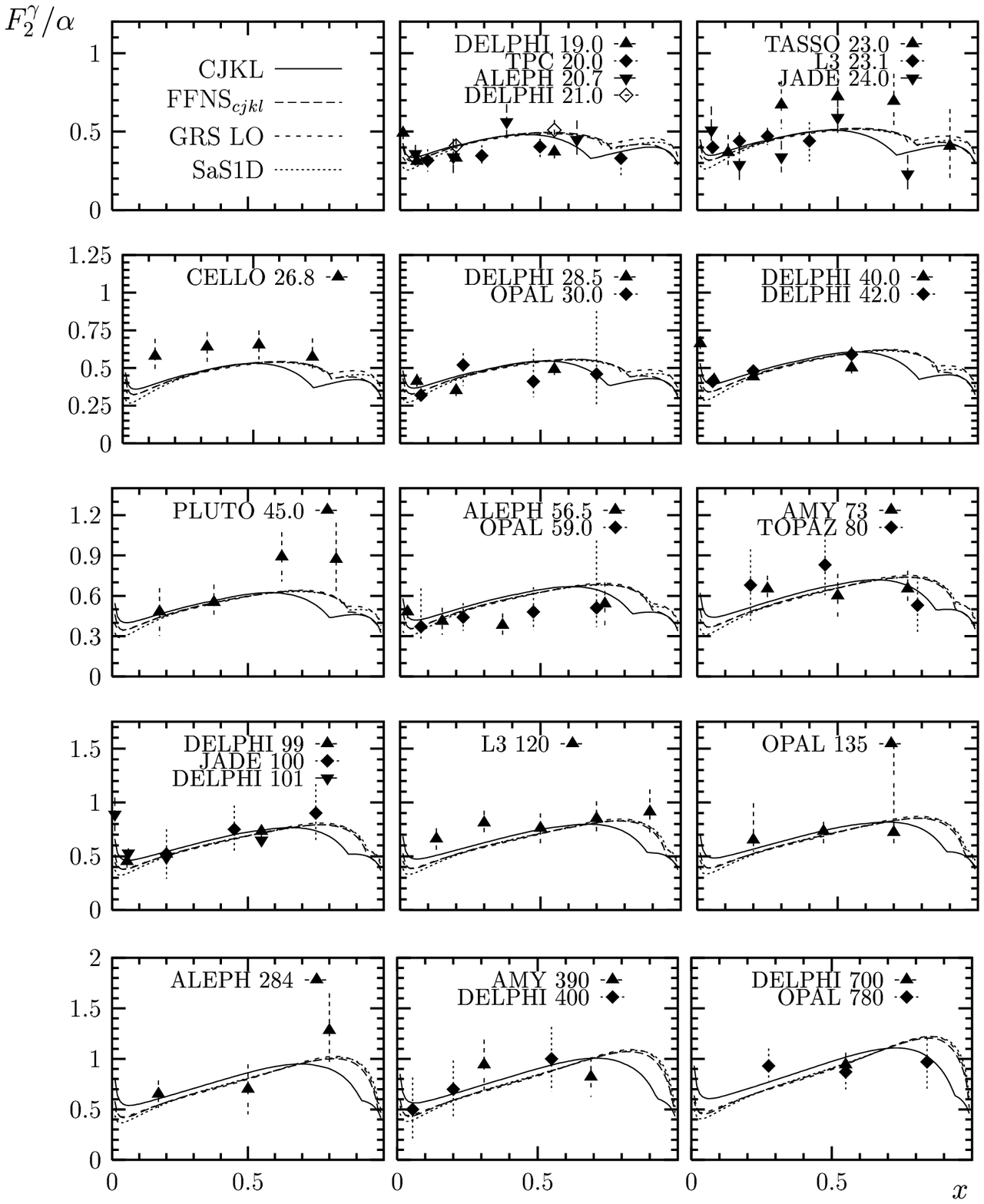}%
\caption{The same as in Fig. \ref{fit2} for a linear scale in $x$.}
\label{fit4}
\end{figure}

\newpage

\begin{figure}
\input{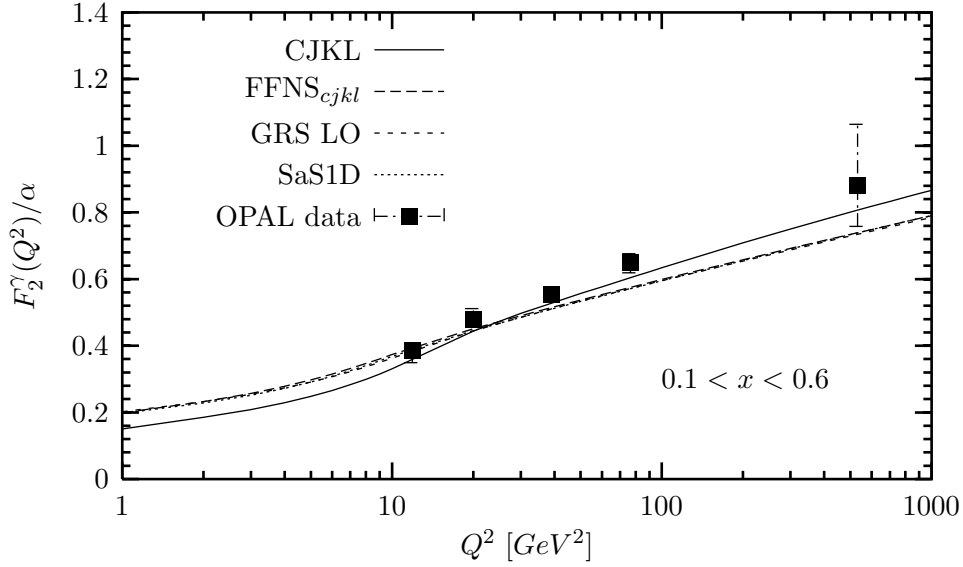}
\caption{Comparison of the recent OPAL data \cite{HQ2} for the
$Q^2$-dependence of the averaged over $0.1<x<0.6$ $F_2^{\gamma}/\alpha $
with the predictions of the CJKL and FFNS$_{cjkl}$ models and GRS LO 
\cite{grs} and SaS1D\cite{sas} parametrizations.}
\label{evol1}
\end{figure}

\begin{figure}[h]
\input{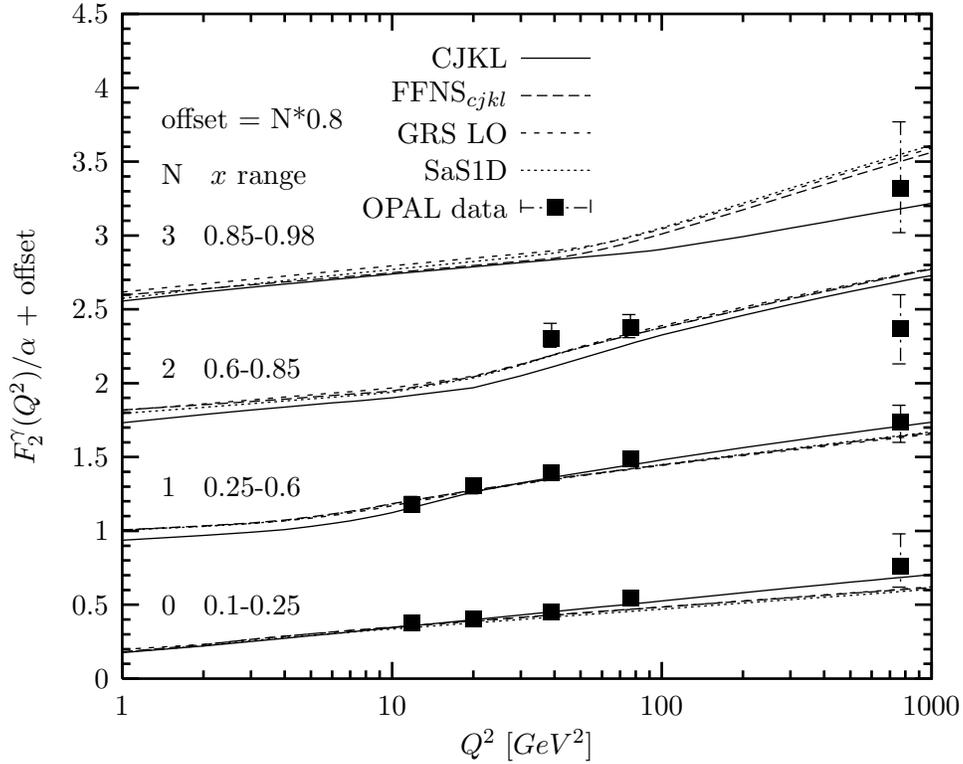}
\caption{
As in Fig.~\ref{evol1} for \fund/$\alpha$, averaged over four different $x$ 
ranges.}
\label{evol2}
\end{figure}

\newpage

\begin{figure}
\begin{center}
\includegraphics[scale=1.0]{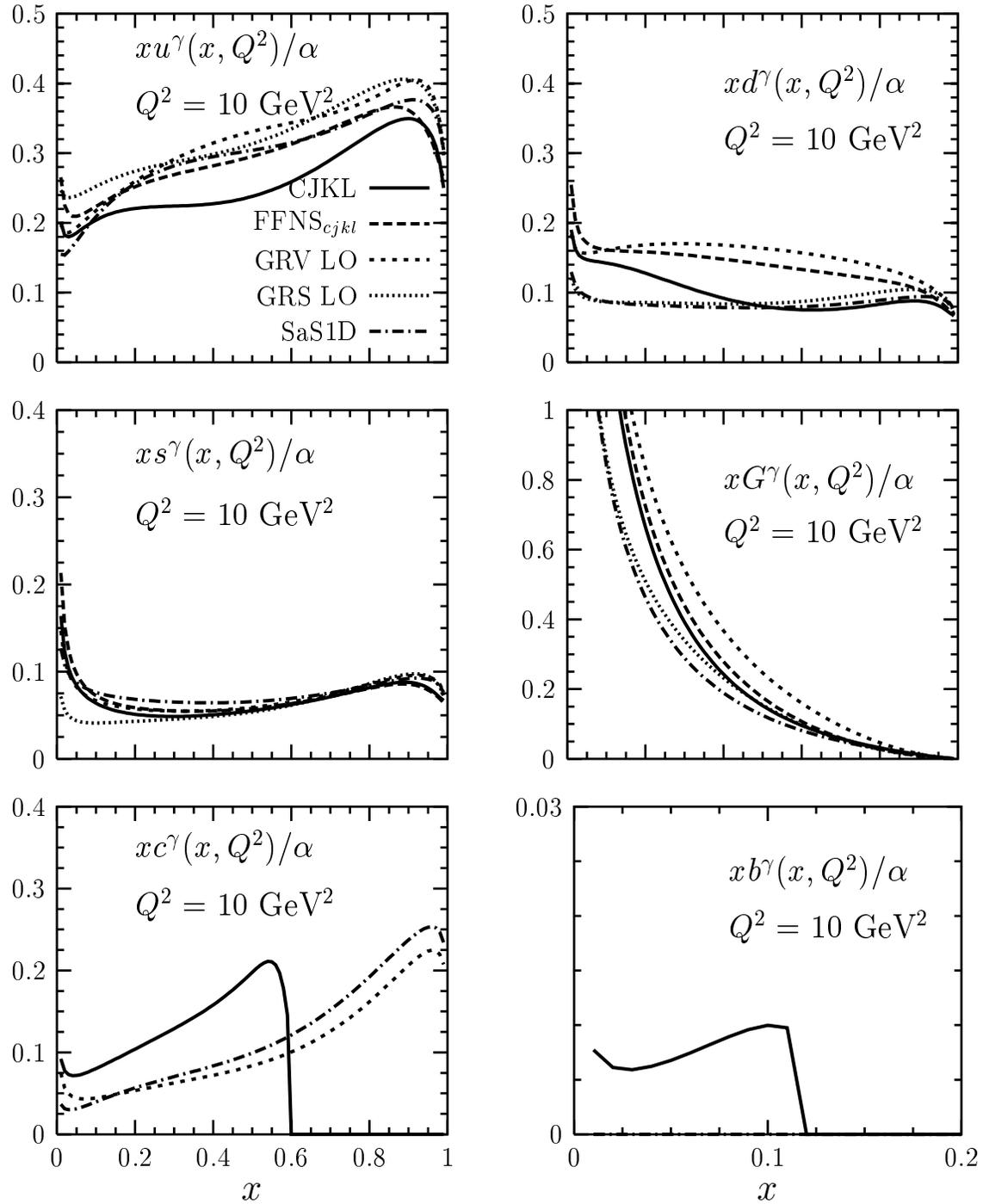}%
\caption{Comparison of the parton densities predicted by various 
parametrizations at $Q^2=10$ GeV$^2$, as a function of $x$. The charm-
and bottom-quark distributions of the CJKL model are compared with the only 
other predictions for heavy-quark densities among the parametrizations used
(given by the GRV LO and SaS1D parametrizations).}
\label{parton}
\end{center}
\end{figure}

\newpage

\begin{figure}
\begin{center}
\includegraphics[scale=1.0]{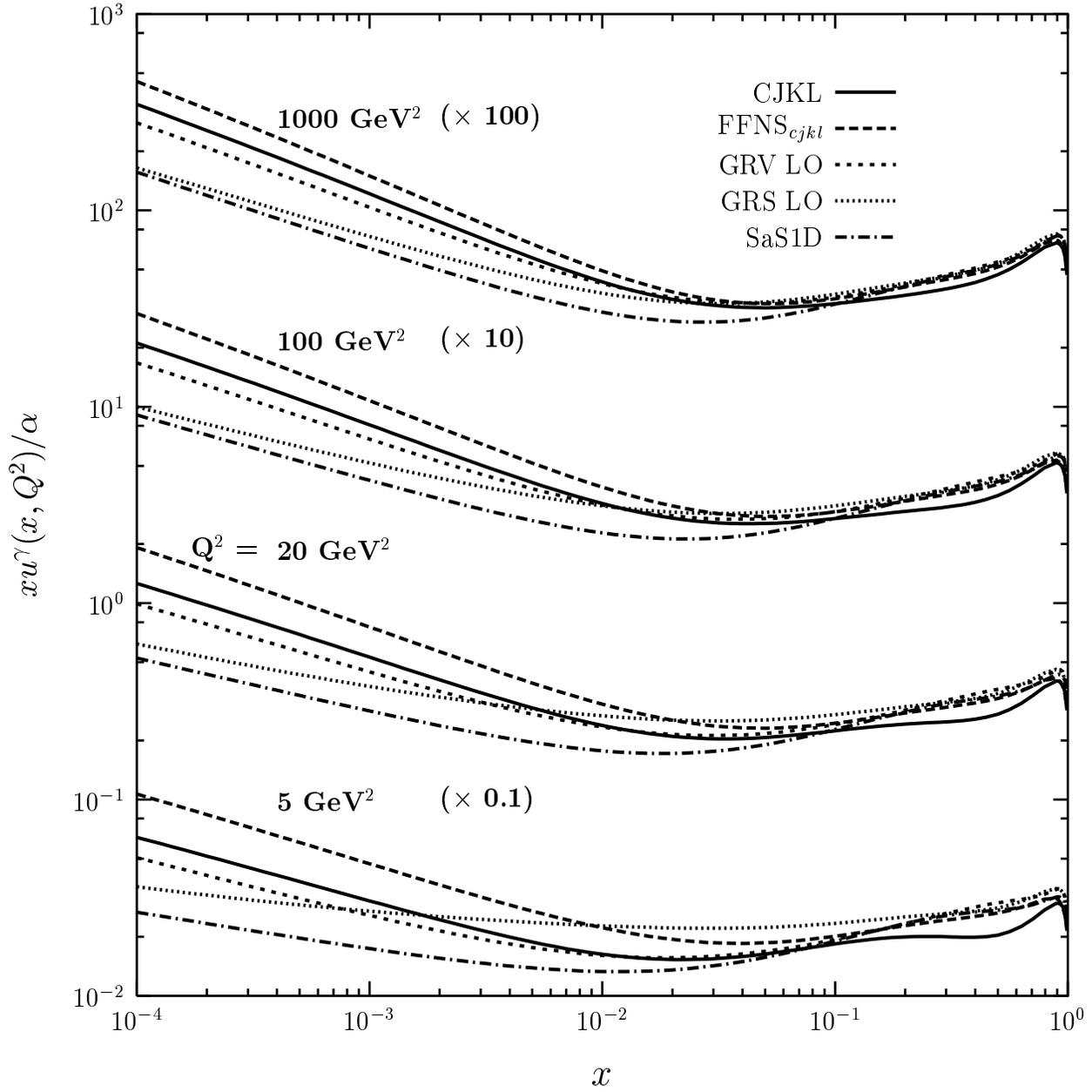}%
\caption{Comparison of the up-quark distribution at four values of $Q^2$ in 
the CJKL and FFNS$_{cjkl}$ models with the GRV LO \cite{grv92}, GRS LO 
\cite{grs} and SaS1D \cite{sas} densities.}
\label{up}
\end{center}
\end{figure}

\vspace{1cm}

\begin{figure}
\begin{center}
\includegraphics[scale=1.0]{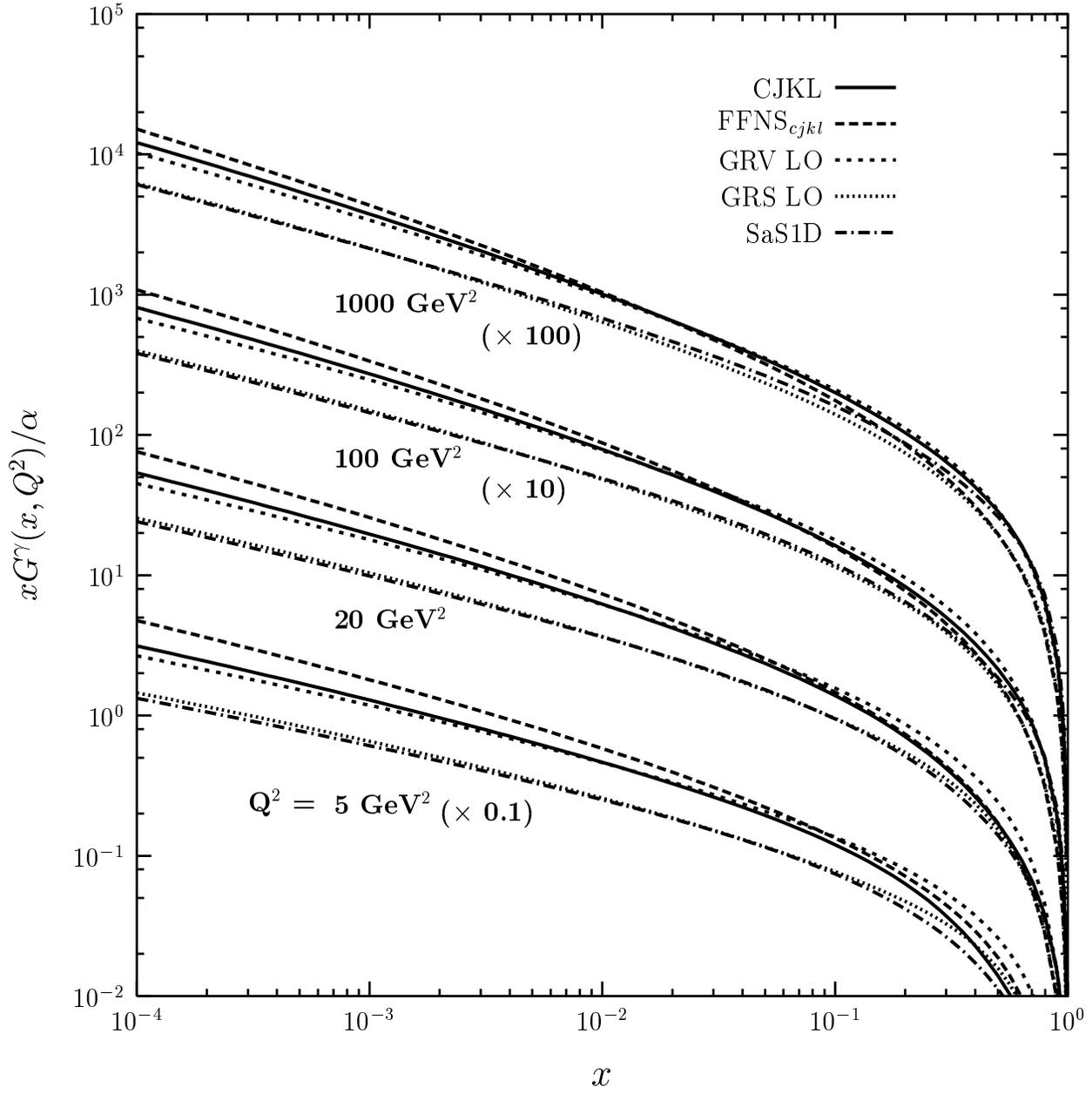}%
\caption{
The same as in Fig.~\ref{up} for the gluon density.}
\label{glu}
\end{center}
\end{figure}

\newpage

\begin{figure}
\begin{center}
\includegraphics[scale=1.0]{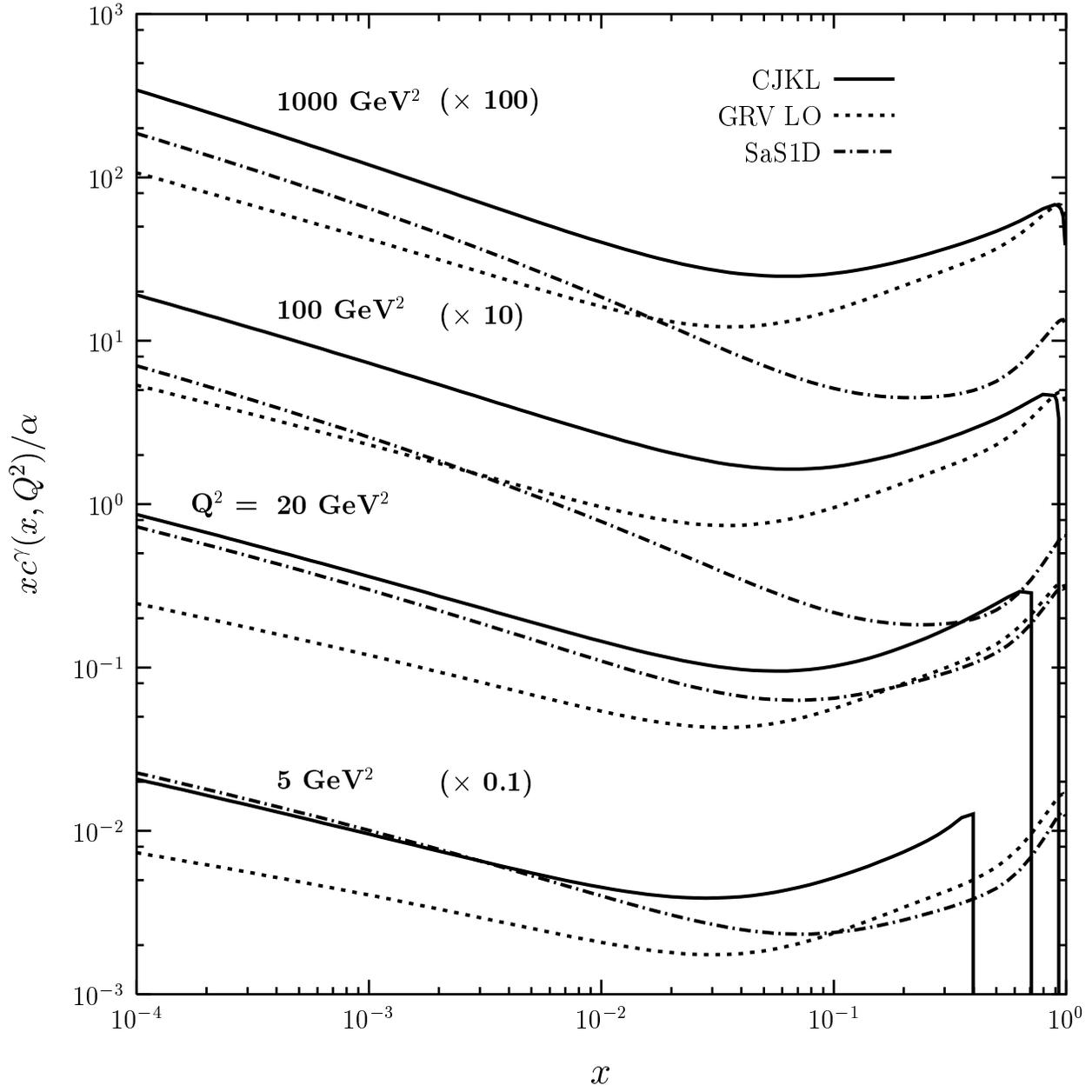}%
\caption{
The same as in Fig.~\ref{up} for the charm-quark  density.}
\label{chm}
\end{center}
\end{figure}

\newpage 

\begin{figure}
\includegraphics[scale=1.0]{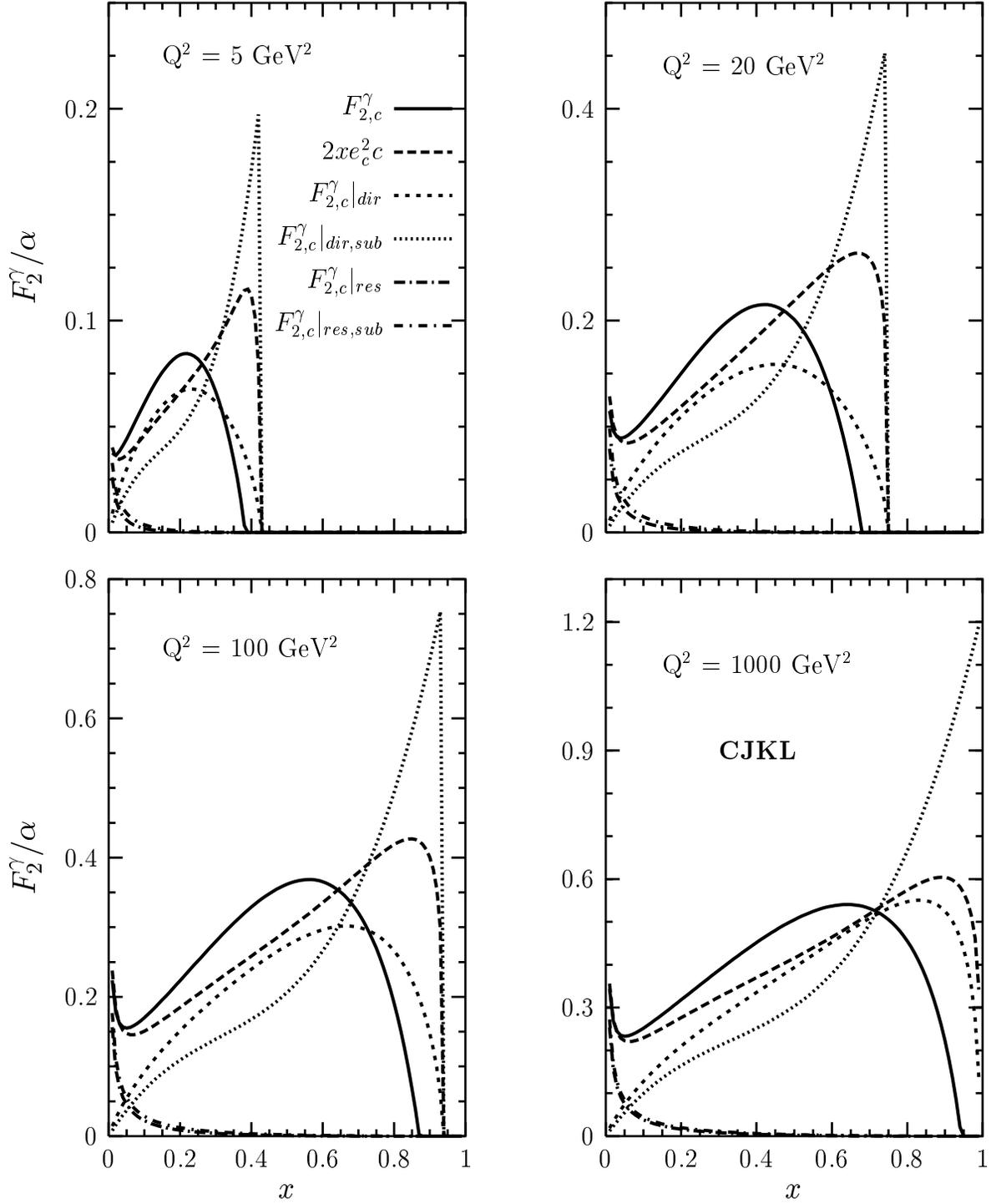}%
\caption{Comparison of various contributions to the photon structure
function $F_{2,c}^{\gamma}(x,Q^2)/\alpha$ in the CJKL model for $Q^2=5,20,100$
and 1000 GeV$^2$.}
\label{acot}
\end{figure}

\clearpage

\begin{figure}[h]
\input{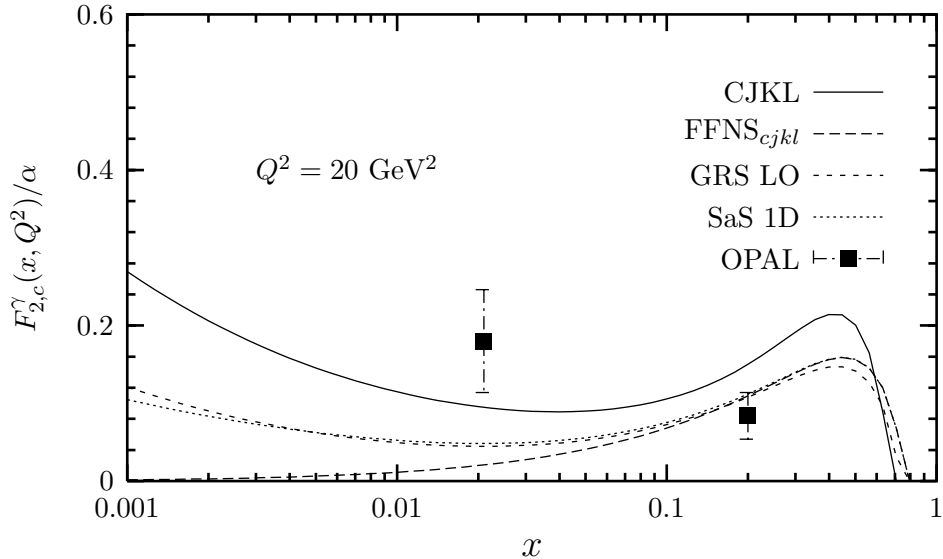}
\caption{Comparison of the structure function $F_{2,c}^{\gamma}(x,Q^2)/\alpha$
calculated in the CJKL, FFNS$_{cjkl}$ models and SaS1D \cite{sas} and GRS LO 
\cite{grs} parametrizations with the OPAL measurement \cite{F2c}.} 
\label{fF2c}
\end{figure}




\end{document}